\documentclass[preprint]{elsarticle}

\usepackage{graphicx}
\usepackage{bm}
\usepackage{hyperref}
\usepackage{color}
\usepackage{amsmath}
\usepackage{amssymb}
\usepackage{amsfonts}

\newcommand{\red}[1]{\textcolor{black}{#1}}

\renewcommand{\d}{\textrm{d}}

\newcommand{\moMS}[1]{10^{#1} M_{\odot}}
\newcommand{\abs}[1]{\left\vert #1\right\vert}

\def\alt{\raise0.3ex\hbox{$\;<$\kern-0.75em\raise-1.1ex\hbox{$\sim\;$}}}
\def\agt{\raise0.3ex\hbox{$\;>$\kern-0.75em\raise-1.1ex\hbox{$\sim\;$}}}

\def\d{{\rm d}}

\newcommand{\GeV}{{\rm GeV}}
\newcommand{\TeV}{{\rm TeV}}

\newcommand{\kpc}{{\rm kpc}}

\newcommand{\cm}{{\rm cm}}

\newcommand{\s}{{\rm s}}

\begin{document}

\title{Dark matter electron anisotropy: a universal upper limit}

 \author[uf,infn,de]{Enrico Borriello}
 \ead{enrico.borriello@desy.de}

 \author[de]{Luca Maccione}
 \ead{luca.maccione@desy.de}

\author[okc]{Alessandro Cuoco}
\ead{cuoco@fysik.su.se}

 \address[uf]{Universit\`a ``Federico II'', Dipartimento di Scienze Fisiche, Via Cintia,  Napoli, Italy}
 \address[infn]{INFN Sezione di Napoli, Via Cintia, Napoli, Italy}
 \address[de]{DESY, Theory Group, Notkestra{\ss}e 85, D-22607 Hamburg, Germany}
\address[okc]{The Oskar Klein Centre for Cosmoparticle Physics, AlbaNova, SE-106 91 Stockholm, Sweden}


\begin{abstract}
We study the dipole anisotropy in the arrival directions of high energy CR
electrons and positrons (CRE) of Dark Matter (DM) origin. We show that this quantity is very weakly 
model dependent and offers a viable criterion to discriminate among CRE  from DM or from local discrete
sources, like e.g.~pulsars.  In particular, we find that the maximum anisotropy which DM can provide is to a very good approximation a universal quantity and, 
as a consequence, if a larger anisotropy is detected, this would constitute a strong evidence
for the presence of astrophysical local discrete CRE sources, whose anisotropy, instead, can be naturally larger than the DM upper limit. 
We further find that the main source of anisotropy from DM is given by the
fluctuation in the number density of DM sub-structures in the vicinity of the observer and  we thus devote special attention to the study of the variance in the sub-structures realization  implementing a dedicated Montecarlo simulation. 
Such scenarios will be  probed in the next years by Fermi-LAT,  providing new hints, or constraints,  about the nature of DM.	
\end{abstract}

\maketitle

\section{Introduction} 
High energy Cosmic Ray
(CR) positrons are promising targets for indirect searches of
Galactic particle Dark Matter (DM) \cite{DMreview}.  The recent results reported by the 
PAMELA \cite{PAMELA} and Fermi collaborations \cite{Abdo:2009zk,Ackermann:2010ij,Ackermann:2011rq} 
on the positron fraction $e^{+}/(e^{+}+ e^{-})$  and on
the $(e^{+}+ e^{-})$ (CRE) spectra in the GeV $\div$ TeV energy range show large discrepancies with standard
astrophysical model predictions and  have indeed raised a large number of  interpretations in terms of  DM. In particular, it has been shown that a good fit of the PAMELA/Fermi data can be achieved with $e^+e^-$ produced by a DM particle of $\sim$TeV mass annihilating or decaying predominantly  via leptonic channels, and several models realizing this scenario have been proposed \cite{Bergstrom:2008gr,Cholis:2008hb,Cirelli:2008pk,ArkaniHamed:2008qn,Bergstrom:2009fa}.
However, interpretations  based on discrete astrophysical extra sources (like e.g.~pulsars, or stochastic local sources)
\cite{Blasi:2009hv,Blasi:2010de,
Serpico:2008te,CRE_interpretation1,DiBernardo:2010is,Blasi:2011fm,Profumo:2008ms,Mertsch:2010fn} have been shown to provide equally good fits to the data (see e.g. \cite{Serpico:2011wg} for an extended critical review of the subject). 

It is, however, very unlikely to distinguish the two scenarios using as observables only the CRE fluxes, even with the larger statistics expected in the future \cite{Pato:2010im}. It is thus mandatory to find other observables accessible to experiments, that are as much model independent as possible and can provide a clear discrimination between a DM dominated scenario and an ``astrophysically'' dominated one. The intrinsic degree of dipole anisotropy in the arrival directions of high energy CREs expected from a DM scenario,
$\delta_{DM}$, is indeed insensitive to many uncertainties, and
constitutes, to a good approximation, a universal characteristics of galactic DM, \red{as we will show in the following}.
The reason why the dipole anisotropy has a very weak dependence on
the many unknowns involved in the problem is, on the one hand, the very short ($\sim 1~\kpc$) electron path above $\sim100~\GeV$  which makes this quantity very local in origin,  on the other hand, the fact that it is a flux ratio (see Eq.~\ref{eq:delta}) so that most of the uncertainties cancel each other in the ratio. 

Furthermore, the anisotropy signal from DM is intrinsically very different from the one due to local discrete sources.  Pulsars are rare (their number in the Galaxy is estimated to be around $10^5$ \cite{FaucherGiguere:2009df}) and powerful objects and can induce very large anisotropies typically dominated by a single or a few nearby objects. On the other hand, the number of galactic DM substructures is ${\cal O}(10^{17})$ and they produce a ``collective'' anisotropy  which is never dominated by a single clump. The flux from a very nearby clump would be always accompanied by the large, dominant and almost isotropic flux from the whole population of clumps, which washes out the single clump anisotropy. 
Therefore, the dipole anisotropy offers a viable criterion to discriminate among CREs produced  by DM or in local sources.

\red{Anisotropies in the DM component have been also studied in gamma-rays (see for example \cite{Cuoco:2010jb,SiegalGaskins:2009ux,Fornasa:2009qh}) which have clearly the advantage of being  independent of the choice of a diffusion models. On the other hand, as we will show, CREs anisotropies are also fairly independent of the propagation setup over a wide range of possible diffusion models and thus offer an interesting complementary anisotropy probe of DM.}

Besides the anisotropy from DM and \emph{local discrete} astrophysical sources, there is also a third source of anisotropy which 
needs to be considered in order to have a complete picture, i.e.  the anisotropy from the large scale distribution of the astrophysical sources considered as a whole. This component, as we will show, gives generally a smaller anisotropy with respect to the first two components above.

On the experimental side, Fermi-LAT recently placed the first upper limits on the integrated dipole anisotropy of the arrival directions of CRE with $E> 60~\GeV$ \cite{Ackermann:2010ip}, and there are prospects for its actual observation after a few years of data taking, if local pulsars contribute significantly to the CRE fluxes above $\sim100~\GeV$ \cite{DiBernardo:2010is}.  Also AMS-02 \cite{AMSweb} is now taking data, but its sensitivity to CRE anisotropy is much lower than the one of Fermi \cite{Pato:2010im}. We will show that if Fermi-LAT or future experiments will find an anisotropy larger than the maximum DM anisotropy we derive here, then a dominant DM contribution to the CRE anisotropy can be excluded in a basically model independent way, pointing instead to local discrete astrophysical CRE sources as the main source of anisotropy. Therefore, the observation of anisotropy at the level within reach by Fermi-LAT in the next years will be able to constrain significantly the flux of CREs possibly contributed by DM annihilations in the Galaxy. On the other hand, to identify pulsars as responsible for a possible anisotropy would require at least a careful analysis of their spectral characteristics and of the direction and intensity of the anisotropy.

This paper is structured as follows: in Section \ref{sec:DMint} we describe how we compute the interstellar CRE density due to astrophysical sources and due to DM annihilations in the smooth halo and in substructures. In section \ref{sec:DManiso} we detail how we simulate the distribution of galactic DM substructures. \red{In section \ref{sec:universal} we compute the total intrinsic DM anisotropy  (i.e. the anisotropy when, ideally, only DM contributes to the total CRE emission) while in section \ref{sec:mixed} we discuss the anisotropy for a mixed scenario in which both DM and standard (non discrete) astrophysical sources  contribute to the CRE flux}. Section \ref{sec:discussion} is finally devoted to our final comments and conclusions.

\section{DM intrinsic electron anisotropy} \label{sec:DMint}
 In the diffusive approach, the dipole anisotropy can be written as
\cite{Berezinsky:book}
%
\begin{equation}
\vec{\delta} = -\frac{3D}{\beta c}\frac{\vec{\nabla} \phi}{\phi}\;,
\label{eq:delta}
\end{equation}
where $D$ is the diffusion coefficient, $\beta c$ and $\phi$ are
the CRE velocity and flux respectively. The total DM contribution
to the $e^+e^-$ fluxes can in general be written as the sum of two components,
$\phi_{DM} = \phi_{h} + \phi_{s}$, where $\phi_{h}$ is the
contribution from the smooth halo while $\phi_{s}$ is the
contribution from the substructures. For each contribution, we have
\begin{eqnarray}
 \phi_{i}(E) &=& \frac{\beta c}{4\pi}\frac{\langle\sigma v\rangle}{2}\left(\frac{\rho_{\odot}}{m_{\chi}}\right)^{2}\int_{V} d^{3}\vec{x}' \int_{E}^{m_{\chi}}dE' \\
\nonumber
&&  \times G(\vec{x}_{S}, E \leftarrow \vec{x}', E')\rho_{i}^{eff}(\vec{x}')^{2}  \frac{dN_\chi}{dE'}(E')
\label{eq:fluxs}
\end{eqnarray}
where $G$ is the Green function associated to the transport
equation \cite{Berezinsky:book}, $\rho_{\odot}$ is the DM density
at the Solar System position and  ${dN_\chi}{/dE'}$ is the
annihilation spectrum into $e^{+}e^{-}$. The term
$\rho_{i}^{eff}(\vec{x}')^{2}$ is defined as
$\left(\rho_{h}(\vec{x}')/\rho_{\odot}\right)^{2}$ in the case of
the DM halo density ($i=h$), while in the case of the
substructures ($i=s$) is written as $\rho_{s}^{eff}(\vec{x}')^{2}
= \sum_{j}\left(\rho_{j}(\vec{x}')/\rho_{\odot}\right)^{2}$, with
the sum running over the substructures and $\rho_{j}$ representing
the DM density of the single substructure.

The large scale anisotropy from astrophysical sources will be estimated
from the interstellar electron density computed numerically with the DRAGON code \cite{dragonwebsite}. 
We will indicate in the following this contribution with the acronym AP. It is worth stressing, however,  that this does not include the anisotropy from discrete sources like pulsars which is in general larger (see \cite{DiBernardo:2010is} and \cite{Blasi:2011fm}) and which is the contribution that we want to isolate in the case of detection of an anisotropy larger than the one provided by DM.

\subsection{Electron propagation}
We solve in the stationary limit $\partial N / \partial t = 0$ the well known diffusion-loss equation \cite{Berezinsky:book}
\begin{equation}
\frac{\partial N}{\partial t} - {\vec{\nabla}}\cdot \left( D(E){\vec{\nabla}}
N\right) - \frac{\partial}{\partial E} \left(b(E)N\right)  =  Q(E,{\vec{x}})\;,
\label{eq:diffusion}
 \end{equation}
where $N$ is the particle number density, $b(E)$ represents energy
losses, $D(E) = D_{0}(E/3~\GeV)^{\alpha}$ is the (spatially
constant) isotropic diffusion coefficient and $Q$ is the source term. Since $|\vec{\delta}|$ is measured at $E> 60~\GeV$ only diffusion and continuous energy losses affect significantly the propagated spectra, so  we can neglect reacceleration and convection. Moreover, at  high energies leptons cannot travel more than a few kpc
\cite{Berezinsky:book}, hence we assume $b(E)= 1.6\times10^{-16} (E/\GeV)^2~\GeV/\s$ corresponding to a magnetic field
and interstellar radiation field  constant over the
relevant propagation region, whose vertical height scale we fix as
$L = 4~\kpc$. As a further consequence,  the effect of boundary
conditions at $E > 60~\GeV$ is
negligible. It can be checked however that changing $L$ in the
range allowed by CR nuclei constraints \cite{Trotta:2010mx} does
not produce a significant effect. Given that the largest effect on the dipole anisotropy is expected to come from the rigidity dependence of the diffusion coefficient, we consider three different models of diffusion: Kolmogorov-like  turbulence (KOL) with
 $\alpha = 0.33$ and $D_{0}=5.8\times 10^{28}~\cm^{2}\s^{-1}$,  Kraichnan-like  turbulence (KRA) with $\alpha = 0.5$ and $D_{0} = 3\times 10^{28}~\cm^{2}\s^{-1}$, and a last one (HA) with $\alpha=0.7$ and $D_{0}=1.28\times10^{28}~\cm^{2}\s^{-1}$. The values of $D_{0}$ are in agreement with CR nuclei observations \cite{DiBernardo:2010is}. 
Further details on how we solve the diffusion equation and derive the related Green functions are given in \ref{appendix:diffeq}.
\begin{figure*}[tbp]
\centering
\includegraphics[width=0.49\textwidth]{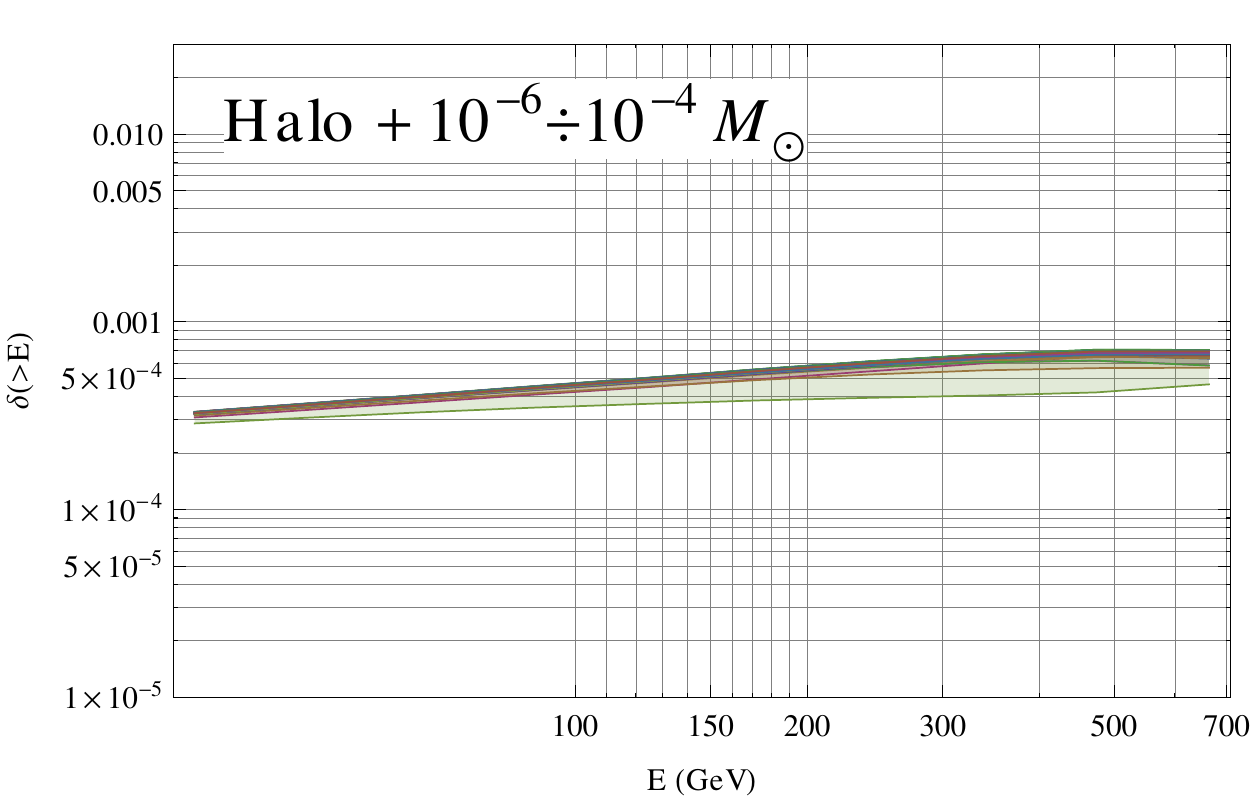}
\includegraphics[width=0.49\textwidth]{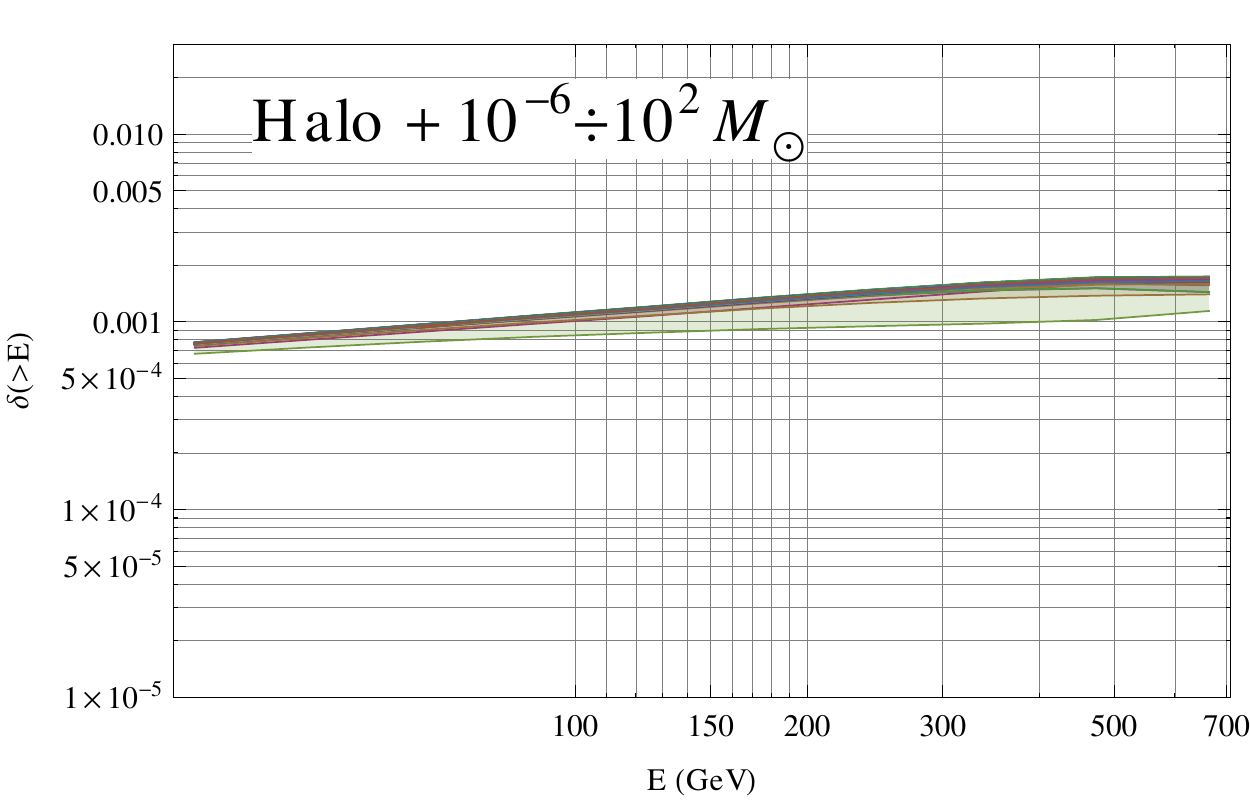}
\includegraphics[width=0.49\textwidth]{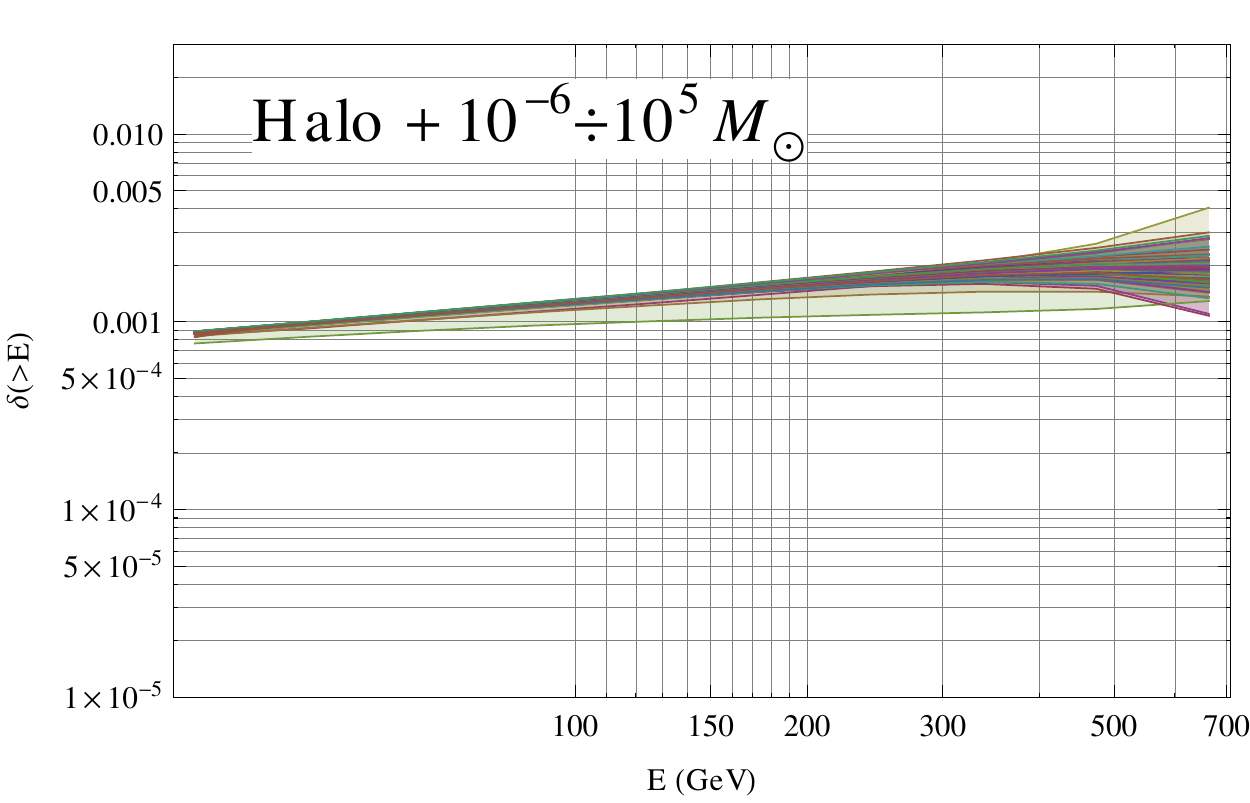}
\includegraphics[width=0.49\textwidth]{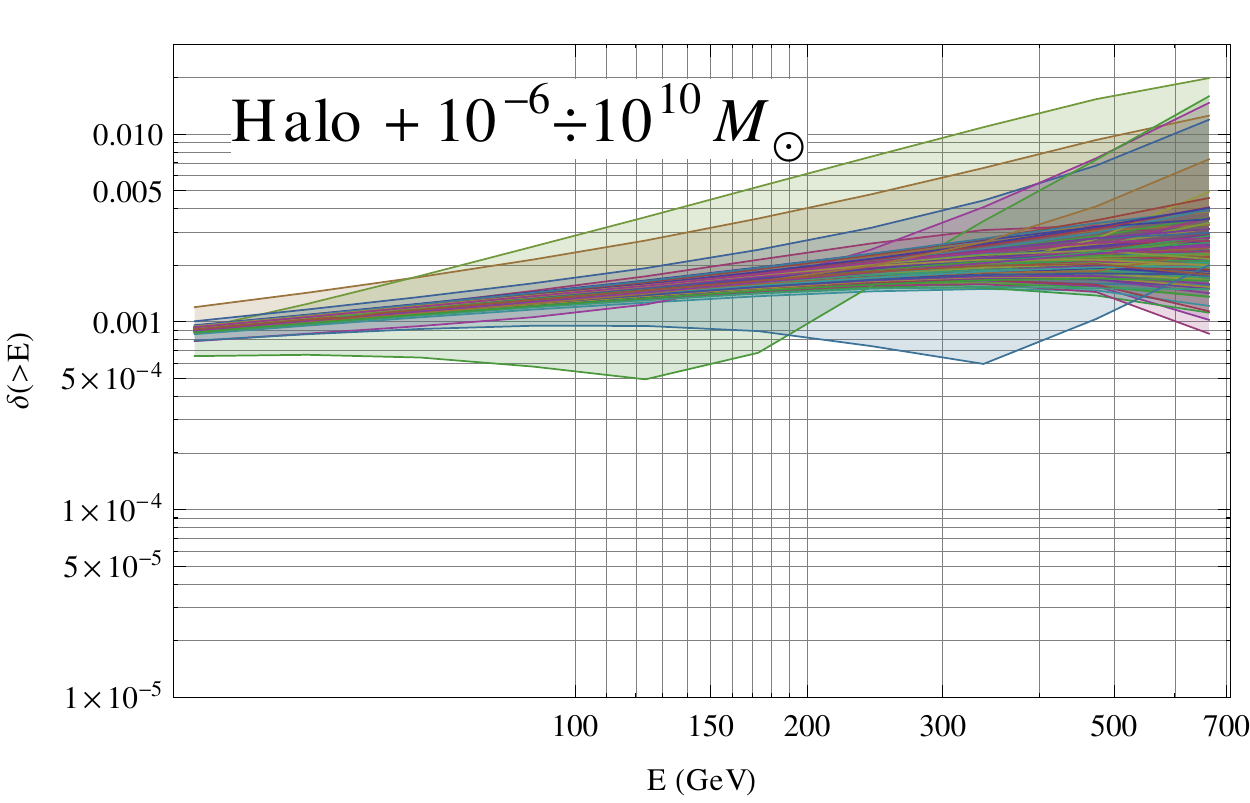}
\caption{DM anisotropy as function of energy contributed by substructures in different mass ranges for 100 different realizations. The plots refers to the case of a 1 TeV DM particle annihilating into $\mu$ pairs, for a NFW distribution and a KRA propagation setup.}
\label{fig:fluctuations}
\end{figure*}

\subsection{Dark Matter distribution}
Beside the diffusion setup another ingredient required to derive the DM anisotropy is its
distribution in the Galaxy.  Current highest resolution N-body simulations (\cite{Diemand:2008in,Springel:2008cc}) find that the DM mass is distributed
roughly equally into a smooth Halo component and into a further clustered part (DM clumps or substructures). 
With respect to the anisotropy, the halo component gives only a mild dipole anisotropy, which for symmetry reasons points toward the Galactic Center.
The clustered component, instead, gives the main contribution to the anisotropy, but  is also more difficult to model. We give below the main points while full details are reported in \ref{appendix:substr}.

$N$-body simulations roughly agree on the mass distribution of substructures, predicting a number density scaling like $m_{cl}^{-2}$ (Via Lactea II \cite{Diemand:2008in}) or $m_{cl}^{-1.9}$ (Aquarius \cite{Springel:2008cc}). How substructures are distributed in the smooth halo is however more uncertain.  We considered the two extreme cases of an unbiased distribution where substructures follow the main halo and an anti-biased case as suggested by the Via Lactea II simulation \cite{Kuhlen:2008aw}. The internal concentration of substructures and the effects of tidal disruption are parameterized as in \cite{Pieri:2009je}. We considered  also a very different set of hypotheses (concentration parametrization taken from \cite{Bullock:1999he} and no tidal effects) finding almost unchanged results, which suggests that internal concentration and tidal forces play a minor role on $\vec{\delta}$. Finally, we chose a clump mass range  $10^{-6} \div 10^{10}~M_{\odot}$, with the upper limit coming from constraints due to disk stability \cite{Ardi:2002ee}. The lower limit is set instead following the most common choice in the literature. Our results however do not depend critically on the lower limit. 
For the spatial distribution of the smooth component and for the DM distribution inside the substructures we consider Navarro-Frenk-White (NFW) \cite{Navarro:1996gj} and Burkert \cite{Burkert:1995yz} profiles. 
\begin{figure}[tbp]
\centering
\includegraphics[width=0.49\textwidth]{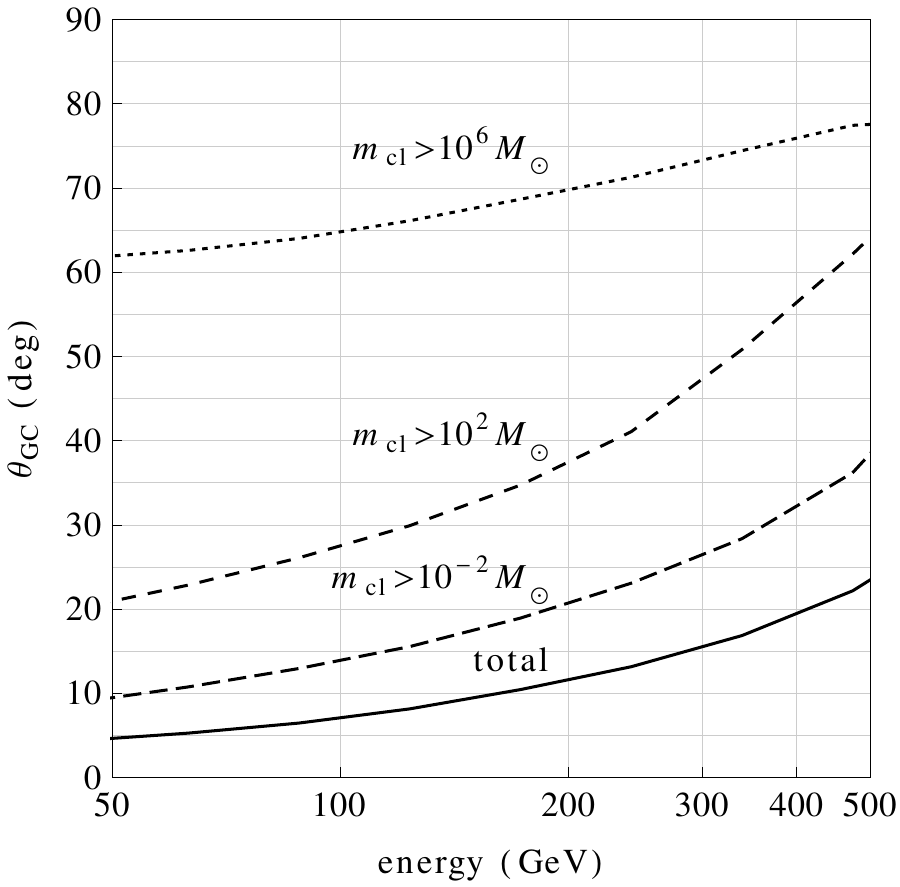}
\caption{Angle between the direction of the DM dipole anisotropy and that of the Galactic Center as function of energy, taking into account substructures in different mass ranges. The lowest curve include the contributions from the smooth halo and the entire substructure distribution, $10^{-6}\div10^{10}M_{\odot}$, averaged over 100 different realizations. The plot refers to the case of a 1 TeV DM particle annihilating into $\mu$ pairs, for a NFW distribution and a KRA propagation setup.} 
\label{fig:angle} 
\end{figure}

\section{DM Anisotropy and Clumps Simulations}
\label{sec:DManiso}

With the explicit solution of the diffusion equation and an analytic form for the distribution of the clumps it is possible to express the average dipole anisotropy resulting from the sum of all the clumps in a completely analytic form which we report in \ref{appendix:mean}. The result shown in \ref{appendix:mean} is valid when the number of clumps contributing to the anisotropy is very large, i.e.~for clumps in the small ($m_{cl}\! < \!\! 10^{2}~M_{\odot}$) mass range, because in this case fluctuations are much smaller than the average.  Clumps of higher mass are instead less abundant and the stochastic fluctuations due to the particular realization of their spatial distribution can give a sizable contribution to the anisotropy which must be taken into account. To this purpose we thus perform explicit Montecarlo simulations of the substructures.

At present, it is computationally prohibitive to simulate the whole population of the ${\cal O}(10^{17})$ substructures lying within the diffusive region. We thus compute analytically the average contribution from substructures with $10^{-6}\!\! <\!\!  m_{cl}/M_{\odot}\!\!  <\!\!  10^{2}$, while we compute explicitly the contribution of each clump with $m_{cl}\!\!>\!\!10^{2}~M_{\odot}$. We sampled the distribution of $m_{cl} \! > \!\! 10^{2}~M_{\odot}$ substructures via a MonteCarlo procedure as described in \cite{Borriello:2008gy}. We produced 100 realizations of substructures with $10^{4}\!\! < \!\!m_{cl}/M_{\odot}\!\!<\!\! 10^{10}$ and 10 realizations with $10^{2} \!\!<\!\! m_{cl}/M_{\odot} \!\!<\!\! 10^{4}$. We then averaged our results over the 100 realizations, having checked that the fluctuations induced by the 10 samples with low masses are much smaller and can be neglected. In total, we computed the contribution of $\cal O$($10^{10}$) substructures for each model.  

According to the analytic computation shown in \ref{appendix:mean}, the anisotropy associated to clumps within a mass decade ($10^n < m_{cl}/M_\odot < 10^{n+1}$, with $-6 \leq n < 2$) is independent of the mass decade (i.e.~it is independent of $n$) to a very good approximation. Therefore, the anisotropy contributed by small clumps results in the sum of several equal contributions. Fluctuations in the distribution of the high mass substructures however break the mass scaling and lead to large variability of $|\vec{\delta}|$. This effect is more relevant at higher energies, because higher energy electrons probe smaller volumes than lower energy ones, thereby being more sensitive to fluctuations in the clump distribution. This is confirmed by the results in Fig.~\ref{fig:fluctuations}  (see also \cite{Pieri:2009je} for an analogous discussion about fluctuations in fluxes) where we show the anisotropy resulting from each of the 100 simulations. In particular, we plot the anisotropy as a function of energy for a DM candidate $\chi$ with $m_\chi=1$ TeV, annihilating into $\mu^+\mu^-$, assuming NFW profile and KRA propagation model, for different clumps mass decades. It is clear from the plot that the main effect of the small mass substructures is to set the average value of the anisotropy, while high mass clumps ($m_{cl}\gtrsim10^{5}~M_{\odot}$) are responsible for fluctuations with respect to the average as large as even one order of magnitude. We remark that, even if the 100 realizations include only clumps in the mass range $10^{4}< m_{cl}/M_{\odot}<10^{10}$, there are fluctuations also for masses  $10^{-6}< m_{cl}/M_{\odot}<10^{4}$, which come from the denominator in the definition of the dipole anisotropy  (that includes the total flux coming from the whole clump sample). These fluctuations are small, in agreement with findings from \cite{Pieri:2009je}.

Finally, we show in Fig.~\ref{fig:angle} the direction with respect to the GC of the DM dipole anisotropy, averaged over our 100 realizations and for different ranges of substructure masses. The plot refers to the case of a 1 TeV DM particle annihilating into $\mu$ pairs, for a NFW distribution and a KRA propagation setup.
While the dipole resulting from high mass clumps only can be pointing in a random direction, the contribution of smaller substructures stabilizes the dipole towards the direction of the Galactic Center. Notice, however, that while at 50 GeV the dipole points within a few degrees to the GC, at 500 GeV the dipole can be pointing up to 25 degrees off.

\begin{figure*}[tbp]
\begin{center}$
\begin{array}{ccc}
\includegraphics[width=0.32\textwidth]{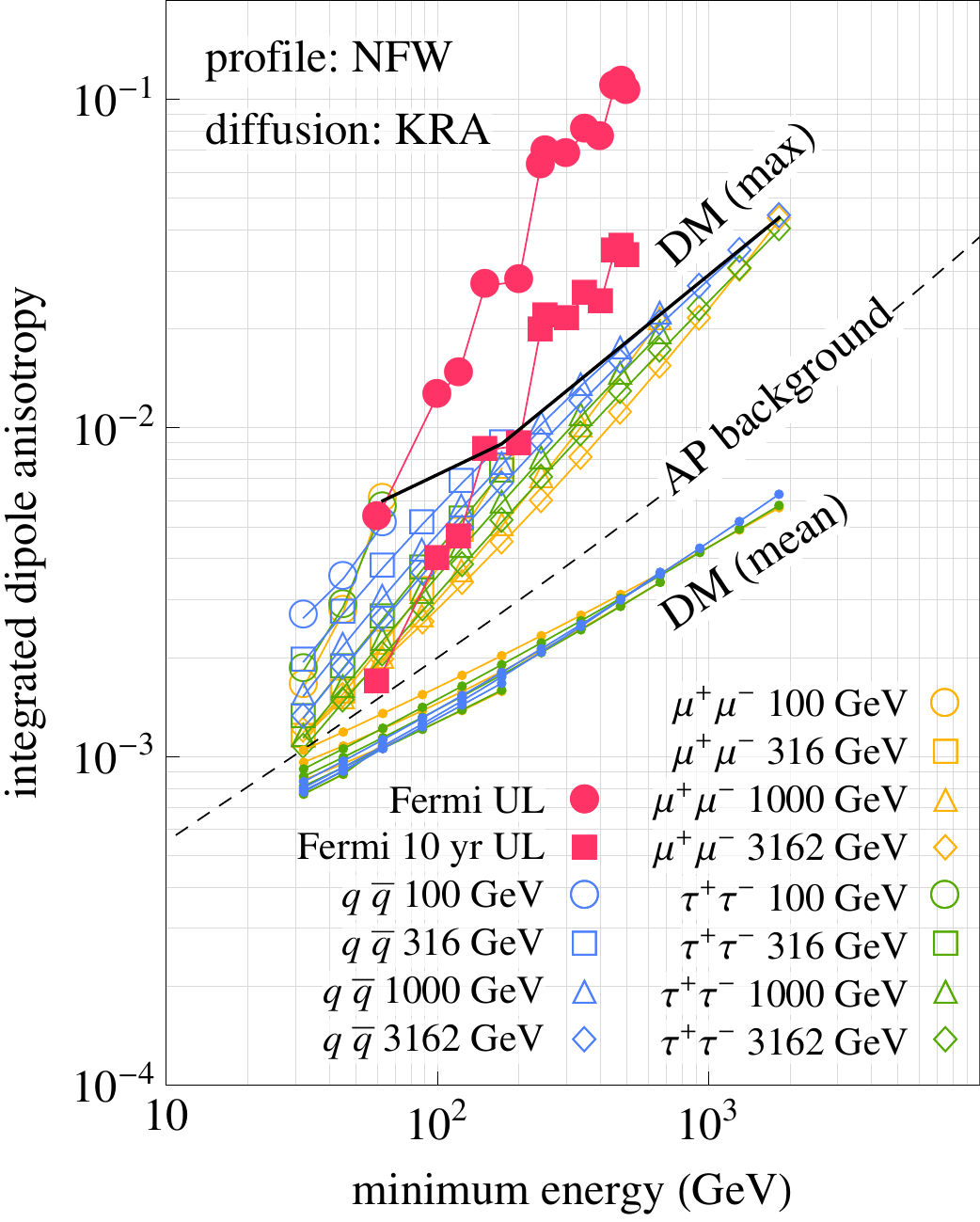} &
\includegraphics[width=0.32\textwidth]{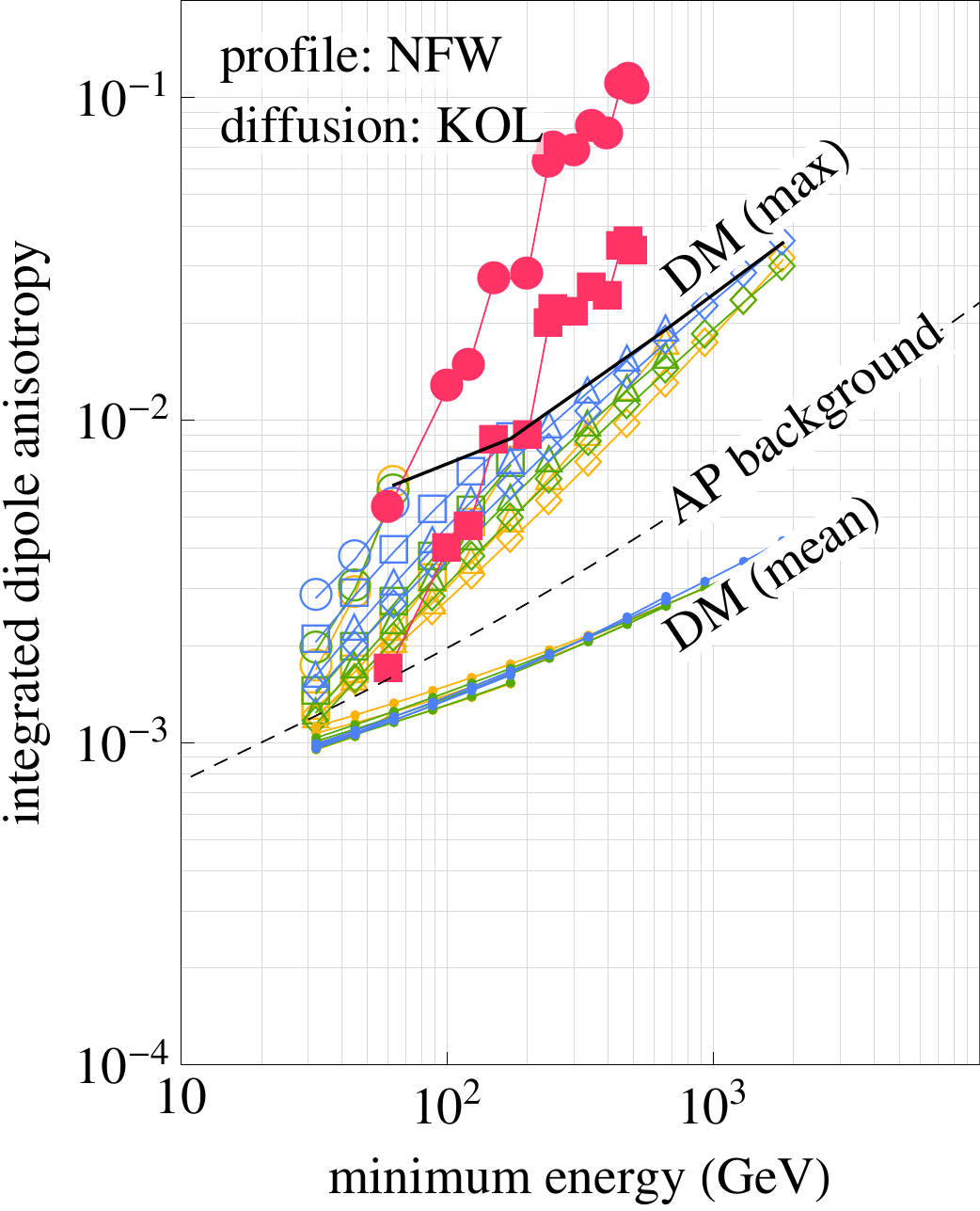} &
\includegraphics[width=0.32\textwidth]{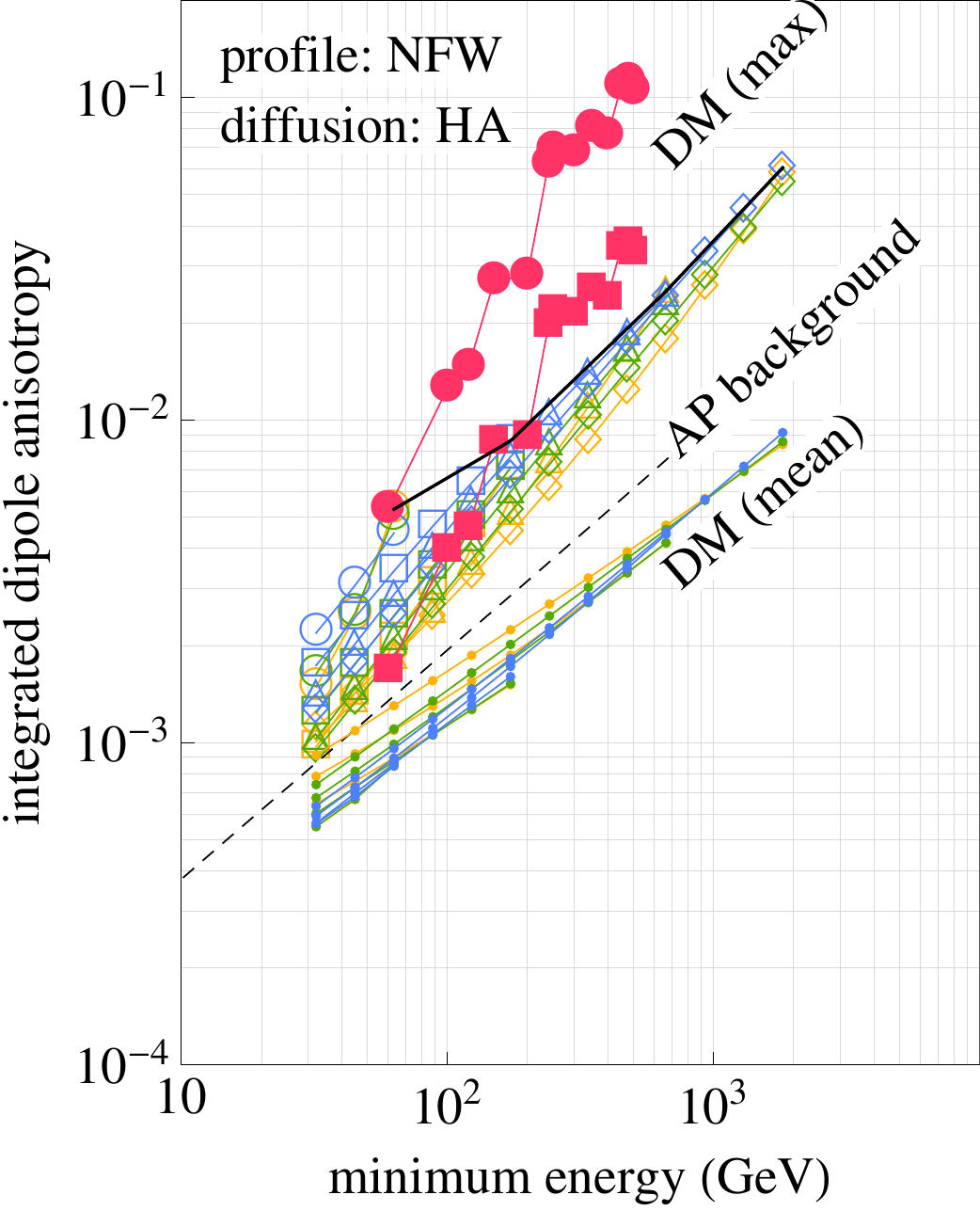}\\
\includegraphics[width=0.32\textwidth]{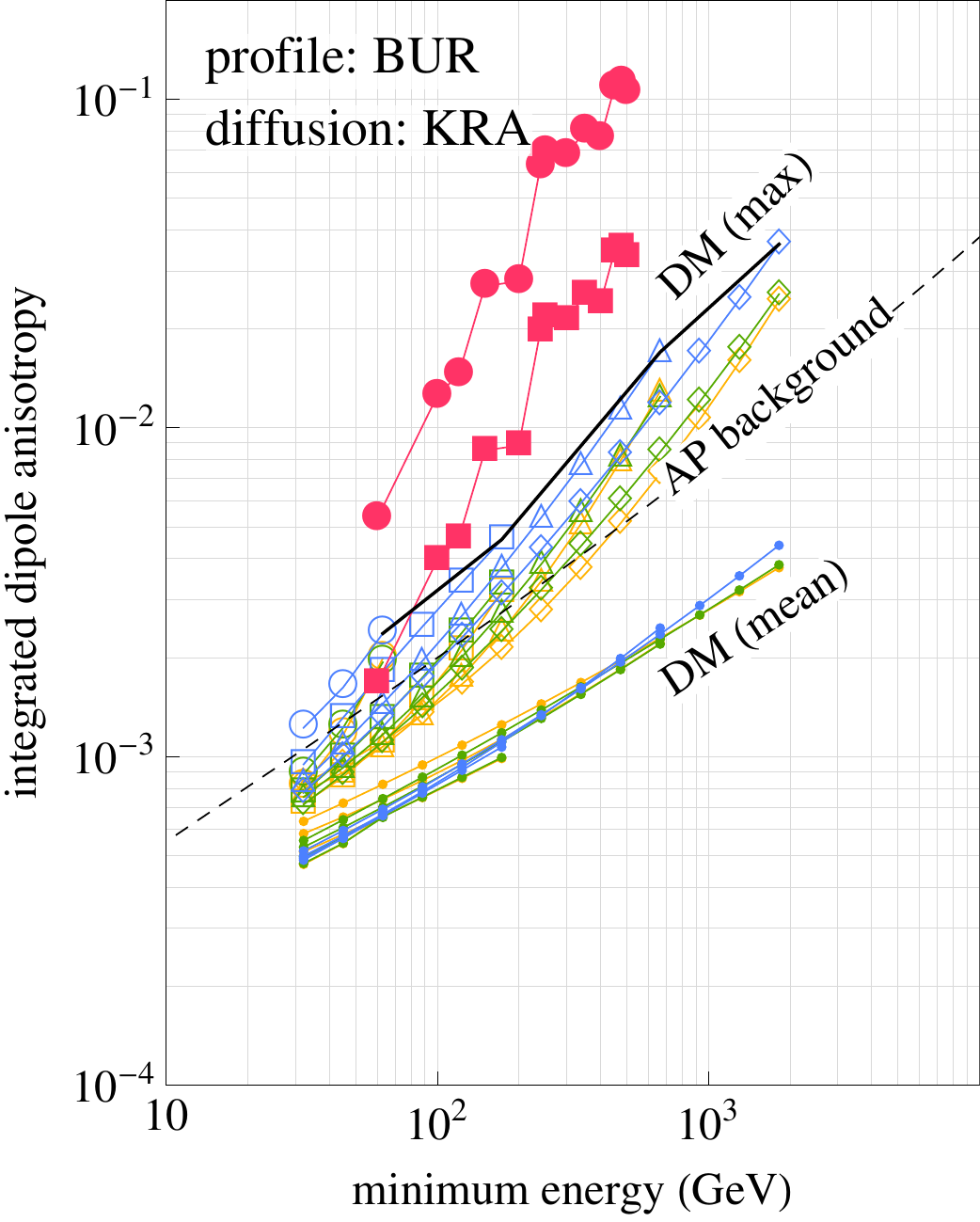} &
\includegraphics[width=0.32\textwidth]{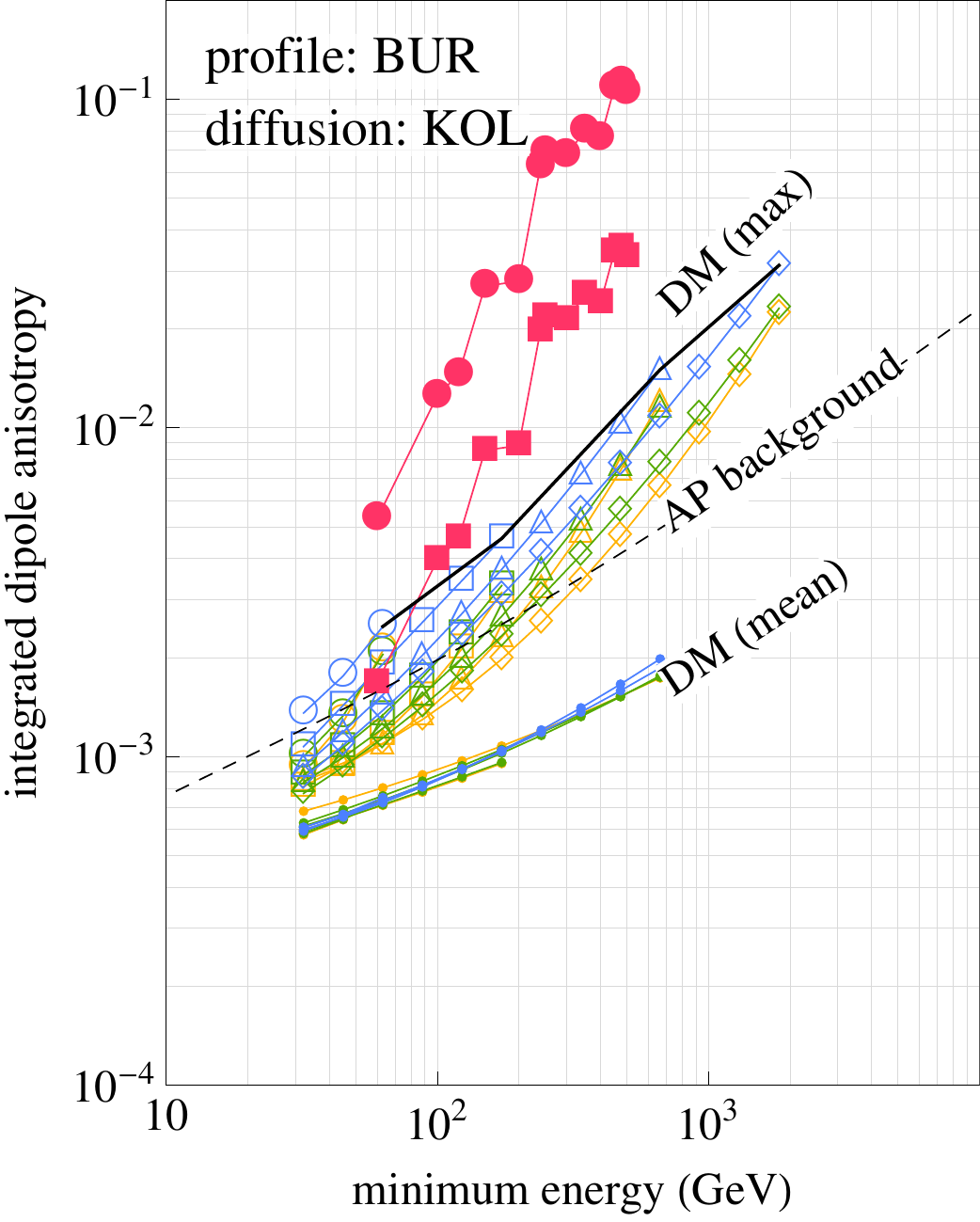} &
\includegraphics[width=0.32\textwidth]{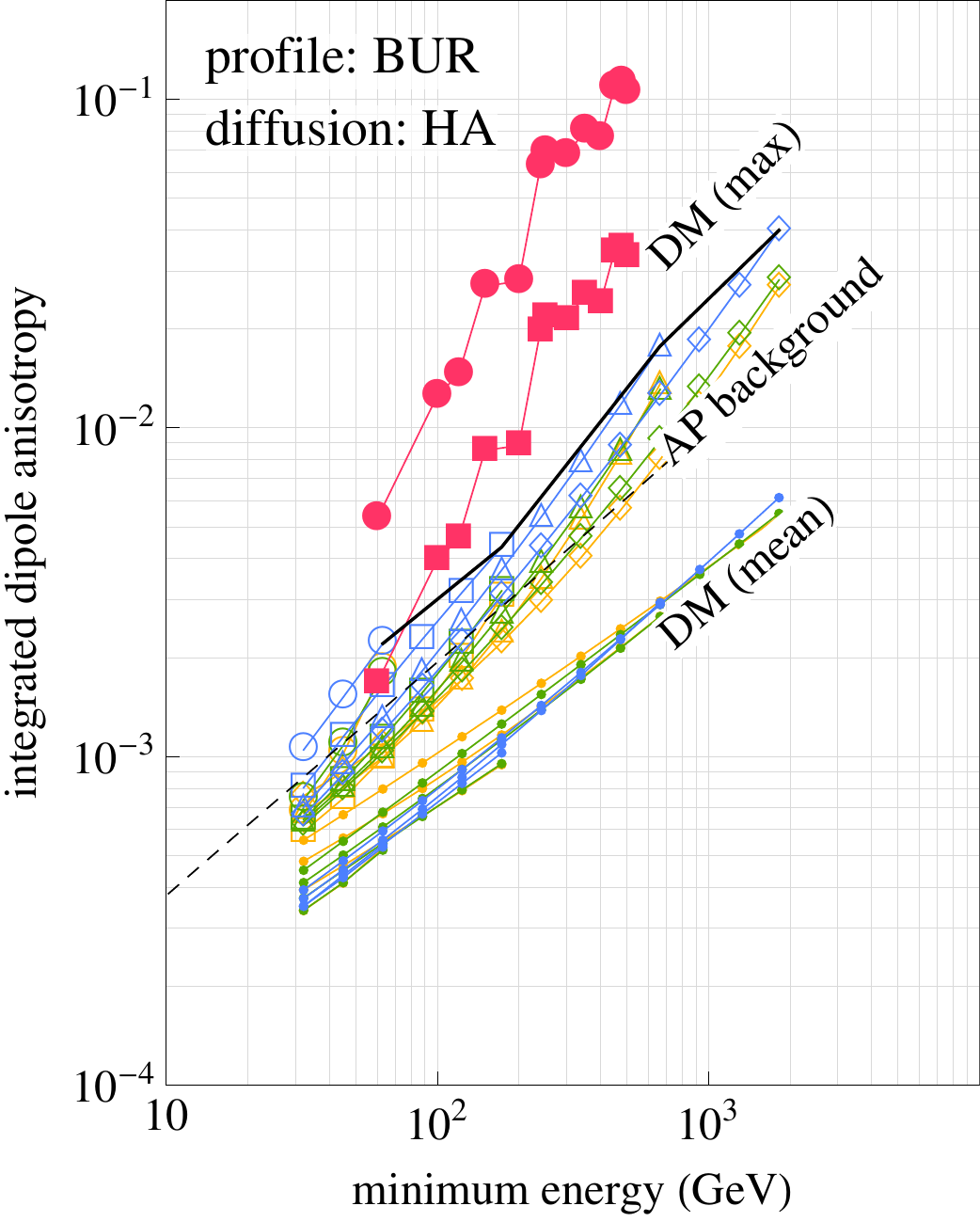}
\end{array}$
\end{center}
\caption{Intrinsic integrated dipole anisotropy of DM, compared to the results for the standard astrophysical background in the associated propagation scenario ($\delta_{AP}$,  black dashed line). Current Fermi upper limits at 95\% CL and the sensitivity expected after 10 years of data taking (actual limits rescaled by a factor $\sqrt{10}$, accounting for the typical $\sqrt{N}$ scaling of sensitivity with statistics -- this represents an optimistic case in which systematics effects do not dominate the measurement) are shown only for reference. The points correspond to different annihilation channels and masses of the DM particle. Both average values and maxima are shown. The thick black line shows the upper envelope of the maxima.} 
\label{fig:universality} 
\end{figure*}

\section{Intrinsic anisotropy upper limits} \label{sec:universal}

The anisotropy signal that can be measured by our observatories must clearly be computed as a suitable composition of the contributions of all the components of CRE fluxes. While we discuss a two-component framework in detail in Section \ref{sec:mixed}, we discuss here the single component scenario in which only DM contributes to the CRE fluxes above 60 GeV. While this assumption is not justified below $\sim200~\GeV$, where data on the positron fraction \cite{PAMELA,Ackermann:2011rq} imply that a possible DM contribution to these fluxes can be at most of the order of 30\%, the single component scenario becomes more realistic at larger energies, where uncertainties on the primary astrophysical CRE component are larger. However, we stick to the use of the single component scenario at all energies in order to retain the simplicity of the discussion and to emphasize the intrinsic properties of the DM anisotropy. For limits below 200 GeV, where a substantial astrophysical component is required, this is, nonetheless, a (overly-)conservative choice, as discussed in more details in the next section.

\begin{figure}[tbp]
\centering
\includegraphics[width=0.48\textwidth]{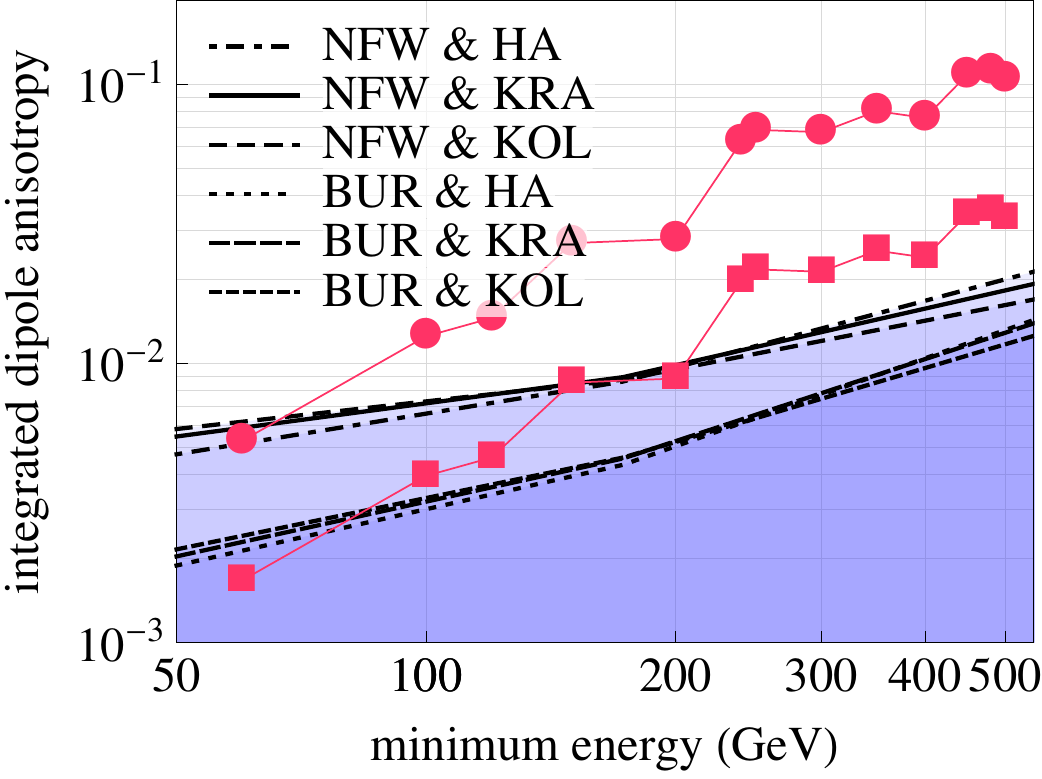}
\caption{Upper limits (UL) on the intrinsic DM anisotropy for all the combinations of DM density profiles and propagation models under scrutiny. Also shown for reference are current Fermi UL and its expected sensitivity after 10 years data taking.
} 
\label{fig:env} 
\end{figure}

The total anisotropy from the smooth Halo and the substructures can be expressed as:
\begin{equation}
\vec{\delta}_{DM} = -\frac{3D(E)}{\beta c}\frac{\vec{\nabla} \phi_{s}(E)+\vec{\nabla}\phi_{h}(E)}{\phi_{h}(E)+\phi_{s}(E)}\;.
\label{eq:DManiso}
\end{equation}
We consider annihilation in $\mu$, $\tau$ and quark pairs, for
values of the DM mass: 100, 316, 1000, and 3162 GeV. \footnote{We compute the injection spectra in these models with the numerical package DarkSUSY \cite{Gondolo:2004sc,DSweb}.}
This basically spans all the possible spectra which can arise from different particle physics model, apart perhaps the case of models where the annihilation is mediated by a light boson as in \cite{ArkaniHamed:2008qn}. 
Even in this case, however, the spectra are only mildly different from the cases considered above and, as we will see, the results are anyway fairly independent of the annihilation channel. Figure \ref{fig:universality} shows the results for the mean and maximum $\delta_{DM}=|\vec{\delta}_{DM}|$ for the considered annihilation channels and DM masses, for our chosen DM density profiles and propagation setups in the unbiased case. The anti-biased case (not shown) gives results smaller by a factor of $\sim 5$ at 500 GeV.  The maximum over our 100 realizations roughly corresponds to a 99\% CL. The figures also show the prediction for the anisotropy of the Astrophysical Background  $\delta_{AP}=|\vec{\delta}_{AP}|$ calculated with DRAGON for the various propagation scenarios considered. 
As already noted, this does not include the contribution to anisotropy from local discrete sources. 
Figure \ref{fig:env} shows, instead, only the envelope of the maxima for the various propagation setups. 
Both the mean and maximum anisotropy increase with energy, as expected from the fact that at higher energies smaller and smaller propagation volumes are probed by the CREs and the role of fluctuations is more relevant.

The main result emerging from the above plots is that $\delta_{DM}$  is almost independent of the detailed characteristics of the DM models and distributions in substructures and, in this sense, $\delta_{DM}$ is a general property of DM. Being a ratio, it is very little sensitive to integrated quantities, like the annihilation spectrum. Moreover, because CREs propagate only a few kpc distance in the Galaxy, $\delta_{DM}$ is also little sensitive to the DM spatial profile, in particular on whether it is peaked or cored. The anisotropy is also not strongly sensitive to the internal concentration of the subhaloes, because diffusion over kpc scales smooths out the effect of a possible cusped over-density region. For these reasons, as we checked, the case of decaying DM gives similar results as the case of annihilating DM.  Finally,  remarkably, also the fluctuations in $\delta_{DM}$ are basically model independent. Figure \ref{fig:env}, in particular, shows that the maximal intrinsic DM anisotropies are almost independent of the diffusion setup, even in the extreme case of HA diffusion, and only slightly change for different DM profiles.

\section{Anisotropy upper limits in mixed astrophysical-Dark Matter scenario}
\label{sec:mixed}

The results described in Section \ref{sec:universal} refer to anisotropy from DM only. We now consider again the role of the background. To fix ideas, we will consider for the AP background an injection spectrum of CRE $dN/dE \propto E^{-2.65}$, which we then propagate with DRAGON using the same propagation setup as for the DM contribution and we normalize such that it accounts for 90\% of the CRE flux at $\sim30~\GeV$, similarly to what done in \cite{DiBernardo:2010is}, where this model was introduced in order to simultaneously fit the CRE fluxes and the positron fraction in a two-component scenario (see \cite{DiBernardo:2010is} for further details). We remark however that our qualitative results do not depend on background assumptions.

If the total flux is given by the contribution of an astrophysical (AP) flux and a DM originated one, $\phi_{TOT} = \phi_{AP} + \phi_{DM}$, the degree of anisotropy is given by a composition of the AP and the DM intrinsic anisotropies:
\begin{equation}
\vec\delta = -\frac{3D}{\beta c}\frac{\vec\nabla\phi_{TOT}}{\phi_{TOT}}\\
           = \frac{\phi_{AP}}{\phi_{TOT}}\vec\delta_{AP} + \frac{\phi_{DM}}{\phi_{TOT}}\vec\delta_{DM}
\label{eq:deltavec}
\end{equation}
Therefore, $\delta=|\vec{\delta}|$ is bounded by
\begin{eqnarray}
 \delta_{min} & = & \abs{
\left(1-\frac{\phi_{DM}}{\phi_{TOT}}\right)\delta_{AP} - \frac{\phi_{DM}}{\phi_{TOT}}\delta_{DM} 
}  \\
\delta_{max} & = & \
\left(1-\frac{\phi_{DM}}{\phi_{TOT}}\right)\delta_{AP} + \frac{\phi_{DM}}{\phi_{TOT}}\delta_{DM} 
\ 
\end{eqnarray}
where, again,  $\delta_{DM}=|\vec{\delta}_{DM}|$ and  $\delta_{AP}=|\vec{\delta}_{AP}|$.

%
%

\begin{figure*}[tbp]
\centering
\includegraphics[width=0.45\textwidth]{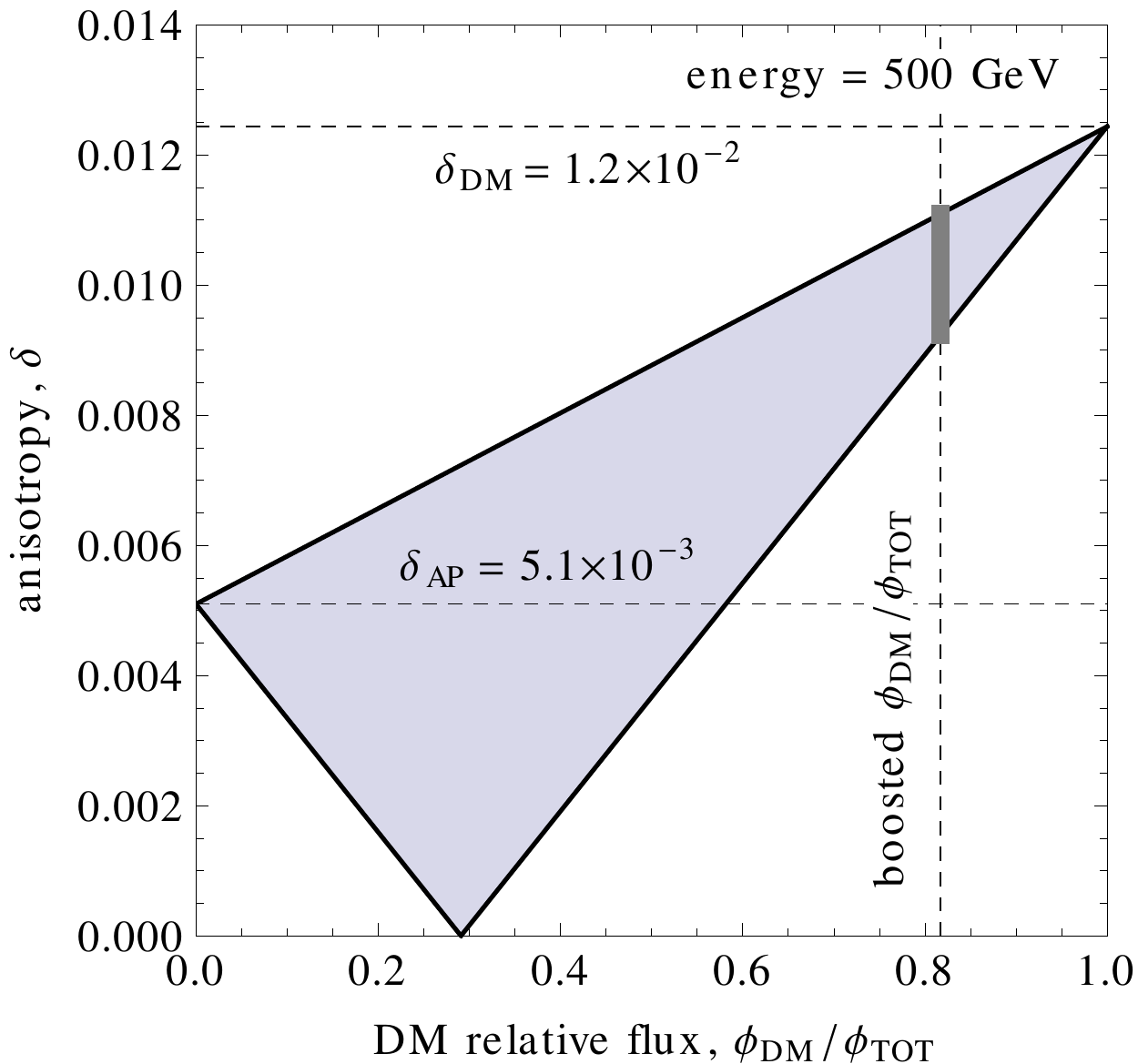} \hspace{2pc}
\includegraphics[width=0.45\textwidth]{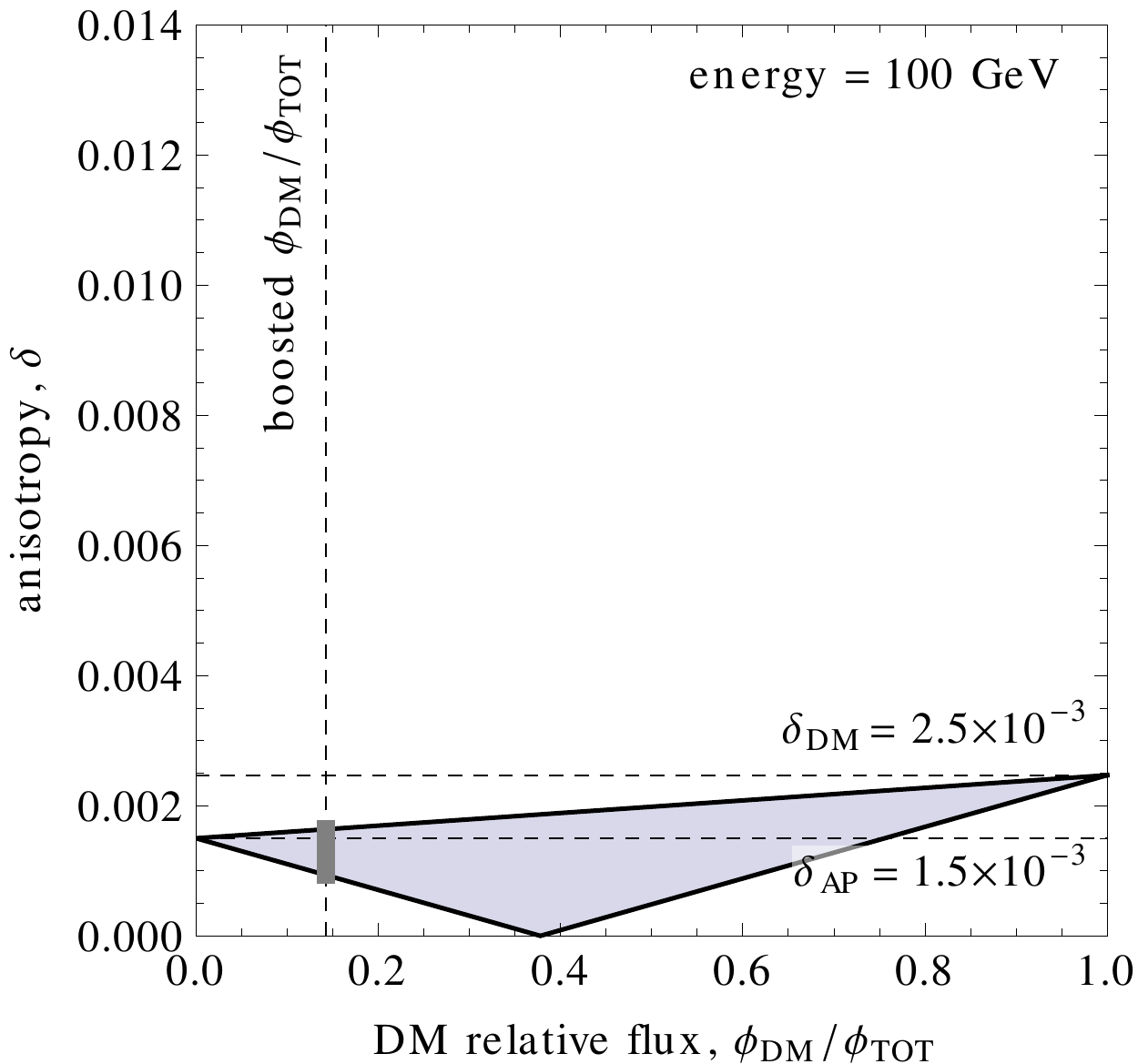} \\
\caption{Allowed region of total CRE dipole anisotropy as a function of the relative DM flux in the mixed (astrophysics + DM) scenario. For a given DM flux the allowed range for $\delta$ is given by the thick gray segment determined by the intersection of the vertical line, corresponding to $\phi_{DM}/\phi_{TOT}$, with the blue shaded triangle.  See text for more details.} 
\label{fig:triangle} 
\end{figure*}

In a specific scenario, $\delta_{max}$ and $\delta_{min}$ are determined by the relative contribution of $\phi_{DM}$ to the total flux.  
The situation is represented for an energy $E=500~\GeV$ by the triangle in Fig.~\ref{fig:triangle} where $\delta_{max}$ and $\delta_{min}$ are plotted as a function of  $x=\phi_{DM}/\phi_{TOT}$. The upper side of the blue shaded triangle represents $\delta_{max}$, while the lower curve shows   $\delta_{min}$.  The shaded region between the $\delta_{max}$ and $\delta_{min}$ curves represents the allowed region which  the total anisotropy $\delta$ can span. 
The lower vertex given by  the value  $x=\phi_{DM}/\phi_{TOT}=\delta_{AP}/(\delta_{AP}+\delta_{DM})$     represents the particular case in which  $\vec\nabla\phi_{DM}$ and $\vec\nabla\phi_{AP}$ are equal and point towards opposite directions and, therefore, $\delta=0$ (see Eq.~\ref{eq:deltavec}).

More specifically the left panel of Fig.~\ref{fig:triangle} shows  the case of total (AP+DM) anisotropy above 500 GeV  for a 3 TeV DM, NFW profile, $\mu^+\mu^-$ annihilation channel and KRA propagation setup. In this case from Fig.~\ref{fig:universality} we can see that  $\delta_{DM}^{max}\sim1.3\times10^{-2}$
while $\delta_{AP}\sim 5\times10^{-3}$ so that the maximum anisotropy is  given by $\delta = \delta_{DM}$ achieved when $\phi_{DM}\gg\phi_{AP}$.
This special case is actually fairly  representative of the general case. From Fig.~\ref{fig:universality}, in fact, it is always true that $\delta_{DM} > \delta_{AP}$. At the same time, the value $\delta_{DM}^{max}\sim 2\times10^{-2}$ at 500 GeV is basically model independent as it can be better seen in Fig.~\ref{fig:env}.

It is also clear that if the DM and AP fluxes are comparable, $\phi_{DM} \sim \phi_{AP}$, the maximum anisotropy will be  \emph{always lower} than the maximum anisotropy in the case in which DM dominates the flux. Indeed, this intermediate situation represents  the most likely scenario, since the combined fits of the Pamela positron fraction and Fermi CREs already strongly disfavor the case in which DM is the dominant flux component, although still allowing a  substantial contribution. 

To check a realistic scenario we consider the case in which DM+AP is required to not exceed the CRE flux measured by Fermi. In this case we find that the CRE flux produced by a 3 TeV DM candidate can have at most a boost factor of $B=470$ with respect to the flux produced with the usual thermal value of the annihilation cross section  $\langle\sigma_A v\rangle=3\times10^{-26}$ cm$^{3}$s$^{-1}$. In this case the DM accounts for a $x=\phi_{DM}/\phi_{TOT}=82\,\%$ of the CRE flux at 500 GeV. 
Given the value $x=0.82$, the actual allowed range for $\delta$ can be determined by the intersection of the vertical line with the triangle as shown in Fig.~\ref{fig:triangle}.
The thick gray segment shows the allowed range for the anisotropy --the actual value being eventually determined by the relative directions of $\vec\nabla\phi_{DM}$ and $\vec\nabla\phi_{AP}$.
A detailed fit of the Pamela and Fermi data by various groups \cite{Cholis:2008hb,Cirelli:2008pk}
gives a  similar or lower DM contribution than the upper limit above, which would give an even smaller range of allowed anisotropy  $\delta$.  
The right panel of Fig.~\ref{fig:triangle} show the same case but for energy above 100 GeV. Above this energy DM accounts for only 14\,\% of the total flux.

It is clear from the discussion presented above that in any case $\delta_{max}\leq \max(\delta_{DM},\delta_{AP})$, and also that for standard astrophysical background and accounting for DM substructures, the maximal DM anisotropy is significantly larger that the AP anisotropy. Therefore, the maximum DM anisotropy $\delta_{DM}^{max}$ constitutes an upper limit to the theoretical CRE anisotropy in such a scenario. Being $\delta_{DM}^{max}$ very little sensitive to the specific DM scenario and to the details of the CRE propagation (see Figs.~\ref{fig:universality} and \ref{fig:env}), this upper limit is very robust. 

If a positive detection of anisotropy will occur in the future, and the anisotropy will be found larger than $\delta_{DM}$, we can then exclude the presence of a substantial DM contribution, and therefore we have to demand $\delta_{AP}>\delta_{DM}$. This would point to a scenario dominated by local, discrete astrophysical sources, such as 
 pulsars, as the main source of high energy CRE, and would strongly constrain the DM contribution to high energy CRE fluxes.
However, this argument does not exclude that a subdominant contribution from DM annihilation in substructures can still be present \cite{Cernuda:2009kk} (the evaluation of this subdominant contribution will depend on the precise characteristics of the model of local astrophysical sources and is beyond the scope of the present work). We remark, moreover, that in a realistic scenario, as the one discussed above, the maximal theoretical anisotropy would be lowered only by about 30\% (see Fig.~\ref{fig:triangle}) with respect to the theoretical maximum we quote in Fig.~\ref{fig:env}. 

On the other hand, if only upper limits will be placed and turn out to be smaller than the maximum anisotropy, constraints on $\phi_{DM}$ will be possibly placed, using e.g.~Fig.~\ref{fig:triangle} (although an improved discussion within a three component scenario also including local astrophysical sources will be needed in that case.). 

\section{Discussion and Conclusions} 
\label{sec:discussion}

\begin{figure*}[tbp]
\centering
\includegraphics[width=0.45\textwidth]{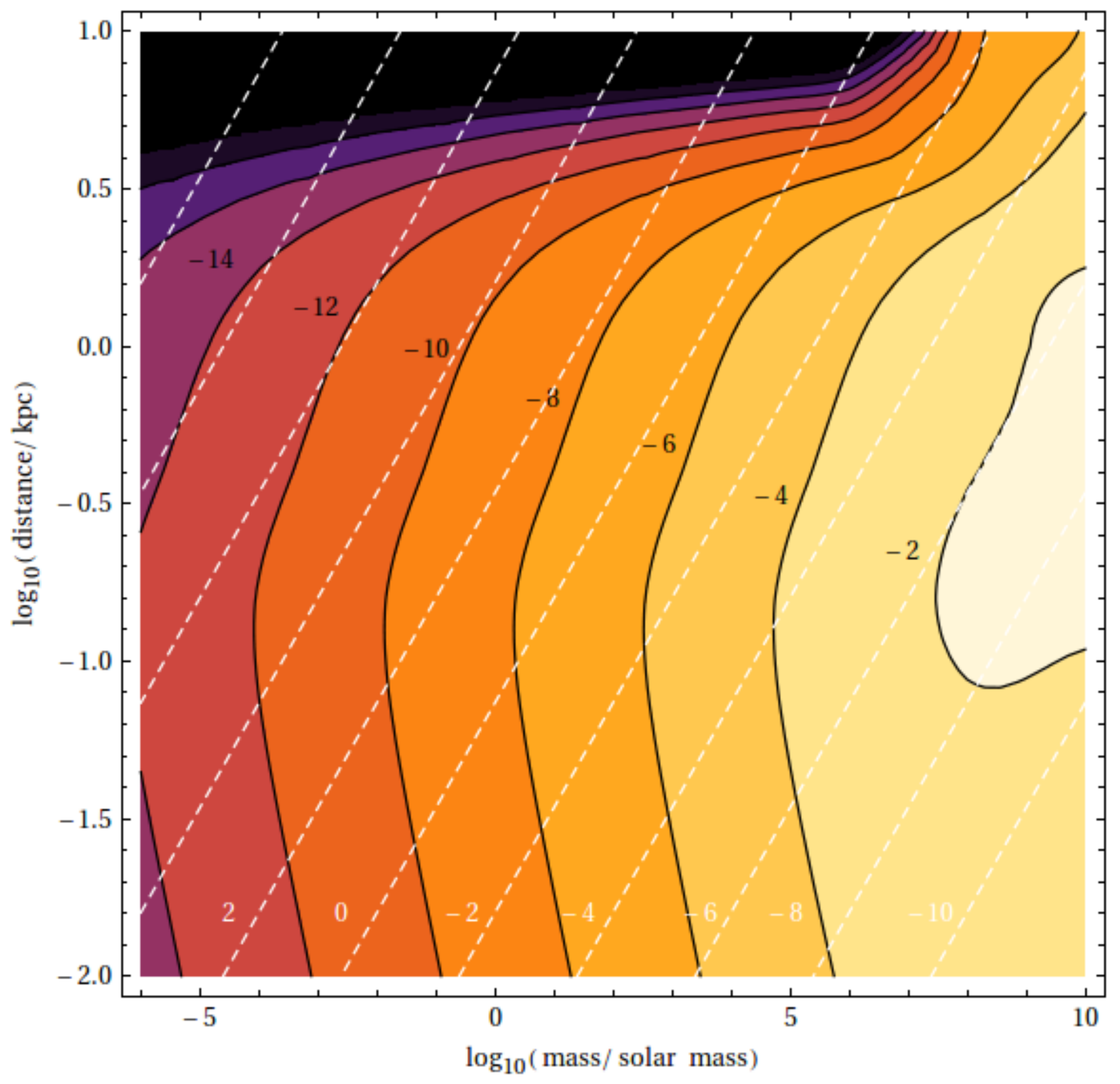} \hspace{2pc}
\includegraphics[width=0.45\textwidth]{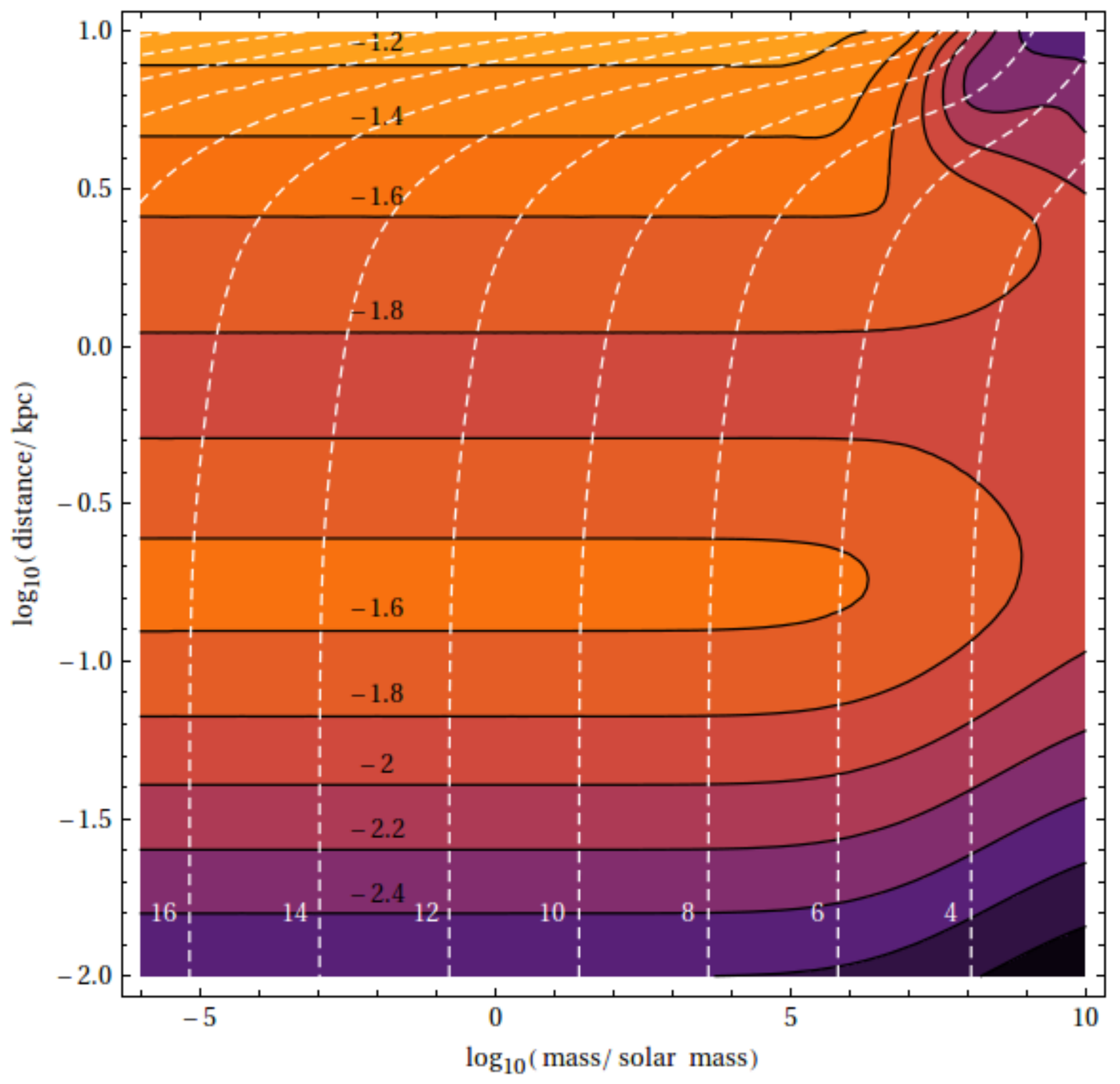}
\caption{\emph{Left}: Black solid contours: anisotropy at 500 GeV due to a single clump
as function of its mass and distance from the Earth when substructures of mass down to $10^{-6}~M_{\odot}$ are included. Grey dashed contours: expected number of clumps of a given mass closer than a given distance to the Earth. Numbers represent the $\log_{10}$ of the related quantities. The plot is drawn for 3 TeV DM fully annihilating into muon pairs, assuming a NFW profile and KRA propagation setup. \emph{Right}: same as left panel, but when no substructures  other than the clump itself (and no main Halo) are included. The white dashed curves represent the log$_{10}$ of the boost factor (with respect to thermal cross section) required for the CRE fluxes to be in agreement with the observed fluxes at 500 GeV.}
\label{fig:singleclump}
\end{figure*}

Our findings result from a MonteCarlo computation of the local distribution of DM substructures and a possible bias of this approach is that we might have missed
configurations whose probability is less than 1\%, in which, e.g., a large mass clump emerges isolated and very close to the Earth. This could in principle produce an anisotropy larger than what we quote as a ``maximum''. We checked, however, that this configuration cannot produce a high degree of anisotropy. Indeed, even in the unlikely case of a $10^{8}~M_{\odot}$ clump at 100~pc from Earth (whose probability is $< 0.1\%$ \cite{Brun:2009aj}) the anisotropy is strongly suppressed by the nearly isotropic flux of the much more abundant  smaller substructures and it is thus always diluted  below  $\max(\delta_{DM})$. 
%
%
%
This feature makes the DM signal intrinsically different from the one expected from pulsars. Indeed, while there might be a close-by, isolated pulsar, that can possibly lead to a large anisotropy \cite{DiBernardo:2010is}, it is not possible to reproduce this configuration with DM. The situation is different also from $\gamma$-rays, where this clump would be a quite bright point source. 

Another possible caveat is that low mass clumps ($m_{cl}\sim 10^{-6}~M_{\odot}$) are so abundant that in principle they can be found within 1 pc from Earth, hence CREs could reach the Earth before diffusing significantly. 
Based on their number density, we expect to find only a few substructures
with mass $10^{-6}~M_{\odot}$ within 1 pc from Earth.
These clumps would look more like point-like sources of $e^+e^-$
rather than like a dipole. Even in this case, however, their point-like flux
both in $e^+e^-$ and $\gamma$-rays would be several orders of magnitude
below the Fermi sensitivity.

These points are illustrated in the left panel of Fig.~\ref{fig:singleclump} where the contours show the anisotropy at 500 GeV of a single clump
as function of its mass and distance from the Earth. 
The plot represents the same case assumed in drawing Fig.~\ref{fig:triangle}: 3 TeV DM fully annihilating into muon pairs and assuming NFW profile and KRA propagation setup. The anisotropy is calculated as in Eq.~\ref{eq:DManiso} with at the numerator the gradient of the flux of the clump and at the denominator the total flux from all the clumps and the smooth Halo so to represent the effective contribution of the clump to the total anisotropy. As it can be seen, the anisotropy has a plateau at about $10^{-2}$ (corresponding to the case where the observer is well inside a very massive clump) which does not exceed the maximum possible anisotropy at the same energy. 

A possible exception to the above scenario is the extreme case in which a massive DM clump is the ``only" relevant CRE source. In fact, any other contribution from other clumps or the smooth halo would dilute the anisotropy of this single clump. We note here that in this scenario, in order for the halo contribution to be negligible with respect to the one of the single clump, one needs to invoke a strong suppression of the annihilation cross section of the DM in the halo much below the thermal value. Moreover, clumps of at least $10^7~M_{\odot}$ are known experimentally to exist in the form of dwarf galaxies \cite{Strigari:2008ib} and they would share the same boost factor as the hypothetical nearby clump, so that their presence would dilute the anisotropy of the single clump anyway. This case is discussed here only because of its extreme geometry and is analyzed in the right panel of Fig.~\ref{fig:singleclump}. The anisotropy is again calculated  as in Eq.~\ref{eq:DManiso} but this time with at the numerator the gradient of the flux of the clump and at the denominator the flux of the clump itself. The needed boost factor with respect to the thermal cross section in order for the clump to produce all the observed CRE flux at 500 GeV is shown as the white, dotted curves. This gives the maximal possible anisotropy and, as we expected, it can exceed the upper limits of Fig.~\ref{fig:env}, being possibly as large as $10^{-1}$.
However, the only clump configurations which would give such a high anisotropy lie in a region of parameter space (small mass or large distance) where unrealistically high boost factors (of the order of $10^6$ or higher) are required. If we restrict the allowed region to the more reasonable case of a nearby massive clump (lower right corner of the plot), again the maximal anisotropy does not exceed $\sim 10^{-2}$ since the anisotropy is somewhat reduced by the fact that the observer is well inside the clump and sees the structure of the clump. We notice that large boost factors (e.g.~via Sommerfeld enhancement) are strongly constrained by analyses of the CMB distortion during the recombination epoch \cite{Galli:2009zc,Slatyer:2009yq,Galli:2011rz}. Further constraints on the boost factor also come for DM annihilation in the core of the Earth \cite{Albuquerque:2011ma}. Also, in the case of a small clump, besides unrealistically high boost factors being required, it would be not justified anymore to not include the contribution of the other equally or more massive clumps, falling back to a configuration of low anisotropy similar to the one of Fig.~\ref{fig:singleclump} (left). For both cases it should be further stressed that, although above $\sim200$ GeV the positron fraction and thus the possible DM  fraction is unconstrained, it is unlikely that DM constitutes 100\% of the CRE flux. Considering the contribution of the astrophysical component will lower the overall anisotropy of the clump although the precise decrease will be dependent on the particular AP model employed.

Another remark concerns the density profiles we considered. 
While $N$-body simulations suggest spiked halo and subhalo matter density
profiles, astrophysical observations of many dwarf spiral galaxies
point to a shallower, Burkert-like density profile
\cite{Salucci:2000ps}. 
Our results are stable under the
relevant change from a spiked to a cored profile. Indeed, high
energy CREs arriving at Earth do not carry information on the DM
distribution in the galactic center, as they propagate only a
few kpc in the interstellar medium. The anisotropy
is not sensitive to the internal concentration of the subhaloes as well,
because diffusion over kpc scales smooths out the effect of a
possible cusped over-density region. For the same reason,
in the case of decaying DM we find similar
results as in the case of annihilating DM.

We also neglected the effects of a possible proper motion of substructures. Indeed, as it was pointed out in \cite{Regis:2009qt} for the case of an isolated substructure, a dynamical treatment would lead to a slightly enhanced dipole anisotropy only for sources moving towards the Solar System. However, while this effect can be relevant for a single clump, it is expected to average away for a population of clumps as considered here.

Finally, a word of caution must be said about our choice of diffusion models. It might in fact be that the local anisotropy observed on Earth can be affected by local magnetic turbulence, which would break the assumption of isotropic and spatially uniform diffusion we used in this work. Hints in this direction may come from the observations of dipole anisotropies and on anisotropies on angular scales of the order of $10^{\circ}\div30^{\circ}$ in the \red{CR hadronic component} at energies $\gtrsim 10~\TeV$ \cite{Abdo:2008aw,Abdo:2008kr,Vernetto:2009xm,Abbasi:2010mf,Toscano:2011dc,Abbasi:2011zk}, \red{as discussed in several works}  \cite{Battaner:2009zf,Salvati:2008dx,Drury:2008ns,Malkov:2010yq,Lazarian:2010sq,Giacinti:2011mz}. 
\red{Anisotropies typically increase as a function of energy on account of the increasing gyro-radius, and  thus the intensity of the anisotropies observed above $\sim 10$ TeV will be correspondingly decreased when rescaled to our energy range  ($\sim$100 GeV - 1 TeV).
On the other hand leptons in this energy range have a much smaller horizon with respect to hadrons  (due to their shorter propagation length) and this typically can increase the anisotropies, since local effects are more important. Thus, it would be  difficult to understand which is the dominant effect and to assess precisely the effects of local turbulence on our results. 
We remark, however, that the observed hadronic dipolar anisotropy seems generally in agreement with the hypothesis of isotropic and homogeneous diffusion \cite{Blasi:2011fm}, while local magnetic turbulence seems to be required mainly to explain the anisotropy at higher multipoles \cite{Giacinti:2011mz}. Since we are considering in this work only dipolar anisotropies, our results are likely less affected by the above effects. }


%

\red{In summary, barring the above caveats, we demonstrate that our results on DM anisotropy are robust with respect to several choices of propagation setup and of DM spatial  distribution and particle model, and we thus propose  to use them as a criterion to reject or at least disfavor a DM dominated scenario in the case of detection of a large anisotropy in high energy CREs. }

\section*{Acknowledgments}
EB and LM gratefully thank T.~Bringmann,  D.~Grasso, M.~N.~Mazziotta, G.~Miele, A.~Mirizzi, P.~D.~Serpico, G.~Sigl,  R.~Tom\'as~Bayo, P.~Ullio and L.~Zhang for stimulating discussions and for reading the draft. EB and LM acknowledge support from the State of Hamburg, through the Collaborative Research program ``Connecting Particles with the Cosmos'' within the framework of the LandesExzellenzInitiative (LEXI).

\appendix

\section{Solution to the diffusion equation}
\label{appendix:diffeq}


Assuming spatially uniform $D$ and $b$ and looking for a stationary solution, Eq.~(\ref{eq:diffusion}) reduces to the form:
\begin{equation}
\label{eqdiffeq}
-D(E) \bigtriangleup \frac{dn}{dE}
-\frac{\partial}{\partial E}\left[b(E)\frac{dn}{dE} \right]
=Q(E,{\bf{x}}) \ .
\end{equation}
We look for the Green function 
$G({\bf{x}},E,{\bf{x}}',E')$ so that the solution of the equation can be written as
\begin{equation}
\label{diffeq}
 \frac{dn}{dE} (E,{\bf{x}})=
\int_E^{m_{\chi}c^2}dE'\int d^3{\bf{x}}'\,G({\bf{x}},E,{\bf{x}}',E')\,Q(E',{\bf{x}}') \ .
\end{equation}
Following  \cite{Baltz:1998,Baltz:2004bb,Colafrancesco:2006,Delahaye:2008}, the Green function $G_F$ for 
the \emph{free} case (i.e.~with no boundary conditions) can be written as
\begin{equation}
 G_F({\bf{x}},E,{\bf{x}}',E')
=\frac{1}{b(E)}
\left(\frac{1}{4\pi\Delta\tau}\right)^{3/2}
\exp\left(-\frac{|\Delta{\bf{x}}|^2}{4\Delta\tau}\right) \ ,
\end{equation}
where we define
\begin{equation}
 \Delta{\bf{x}}={\bf{x}}-{\bf{x}}'
\qquad
\textrm{and}
\qquad
 \Delta\tau=\int_E^{E'}\frac{D(\epsilon)}{b(\epsilon)}d\epsilon \ .
\end{equation}
Therefore,
 \begin{equation}
 \frac{dn}{dE}({\bf{x}},E)=\frac{1}{b(E)}\int_{E}^{m_{\chi}c^2}dE'
 \frac{\langle\sigma_A v\rangle}{2}\left(\frac{\rho_{eff}({\bf{x}},E,E')}{m_{\chi}}\right)^2
 \frac{dN}{dE'} \ ,
 \end{equation}
where $\rho_{eff}$ is such that
\begin{equation}
 \rho_{eff}^2({\bf{x}},E,E')=(4\pi\Delta\tau)^{-3/2}\int d^3 {\bf{x}}' 
\rho^2({\bf{x}}')\exp\left( -\frac{|\Delta {\bf{x}}|^2}{4\Delta\tau} \right) \ .
\end{equation}
In the case of a clump
\begin{equation}
 \rho({\bf{x}}')=\rho_0 f(|{\bf{x}}'-{\bf{x}}_{cl}|/r_0)\ ,
\end{equation}
with ${\bf x}_{cl}$ being the position of the center of the clump.

By changing variables to spherical coordinates local to the clump's center, and using $\xi=|{\bf{x}}'-{\bf{x}}_{cl}|/r_0$, we end up with \footnote{For a NFW profile  $x_{-2}=1$, while 
for a Burkert profile $x_{-2}=(1-\sqrt{26/27})^{1/3}+(1+\sqrt{26/27})^{1/3}\simeq1.52$. } 
\begin{eqnarray}
\label{rhoeff2}
\rho_{eff}^2({\bf{x}},E,E')= 
\rho_0^2\frac{4\pi r_0^3}{(4\pi\Delta\tau)^{3/2}} \frac{2\Delta\tau}{r_0 |\bf{x}-{\bf{x}}_{cl}|}
\exp\left(-\frac{|{\bf{x}}-{\bf{x}}_{cl}|^2}{4\Delta\tau}\right) \nonumber  \\ 
\int_0^{x_{-2} c_{cl}} \xi f^2(\xi)\exp\left(-\frac{r_0^2}{4\Delta\tau}\xi^2\right)\sinh\left(\frac{r_0|{\bf{x}}-{\bf{x}}_{cl}|}{2\Delta\tau}\xi\right)d\xi \ . \nonumber  \\
\end{eqnarray}

An interesting limit is given by the situation in which the observer is well outside the clump. In this approximation  we have
 $|\Delta{\bf{x}}|^2= |{\bf{x}}-{\bf{x}}'|^2  \simeq |({\bf{x}}-{\bf{x}}_{cl})|^2$ and the above expression simplifies  as 
\begin{equation}
 \rho_{eff}^2 \simeq \rho_{cl}^2 \frac{4\pi r_{cl}^3}{(4\pi\Delta\tau)^{3/2}} 
 \exp\left(-\frac{s^2}{4\Delta\tau}\right)
 \int_0^{x_{-2} c_{cl}} \xi^2 f(\xi) d\xi \ .
\end{equation}
where 
$s=|\textbf{x}-\textbf{x}_{cl}|$.


The decaying DM case is completely analogous. We simply have to substitute $\langle\sigma_A v\rangle/2$ with $\Gamma$ and $\rho^2_{eff}$ with a $\rho_{eff}$ defined in terms of $\rho(\textbf{x}')$.

\section{Distribution of substructures}
\label{appendix:substr}

We assume that the number density of subhaloes scales with mass and position according to 
\begin{equation}
\label{dndmc}
\frac{{\d} n_{{cl}}}{{\d} m_{{cl}}}(m_{{cl}},\textbf{x})= A
\left( \frac{m_{{cl}}}{M_{\odot}} \right)^{-{\alpha}} g_{cl}\left(\frac{r}{r_{0}} \right) \ ,
\end{equation}
where $g_{cl}(r/r_0)$ is a dimensionless function parameterizing the spatial distribution of the substructures; $r_0$ is a scale parameter for the entire DM distribution; $A$ is a dimensional normalization constant. 
The effect of  tidal disruption of clumps near the galactic center is included \textit{a posteriori} in our Monte Carlo according to the Roche criterion (see e.g.~\cite{Pieri:2009je}), but not in the following calculations which are normalized  to clumps with mass $ \alt 10^7  M_{\odot}$ and are thus unaffected by the small number of tidally disrupted clumps. Similarly we only use $\alpha=2$, because with a minimum mass scale of $\moMS{-6}$, a mass index of 2.0 or 1.9 produces only a minor change in the results.

From Eq.~(\ref{dndmc}), the total mass and number of  DM clumps of
mass between $m_1$ and $m_2$ can be written as
\begin{eqnarray}
M(m_1,m_2) &=& \int {\d}^3\textbf{x} \int_{m_1}^{m_2} m_{{cl}}\frac{{\d} n_{{cl}}}{{\d} m_{{cl}}}(m_{{cl}},\textbf{x}) {\d}m_{{cl}}\qquad\\ 
N(m_1,m_2) &=& \int {\d}^3\textbf{x} \int_{m_1}^{m_2} \, \frac{{\d} n_{{cl}}}{{\d} m_{{cl}}}(m_{{cl}},\textbf{x}){\d}m_{{cl}}  \  .
\end{eqnarray}
According to \textit{Via Lactea II} \cite{Diemand:2008in} we impose the condition that  the mass contributed by clumps of mass between $10^{7}M_{\odot}$ and $10^{10}M_{\odot}$ is 10\% of the total mass of the Milky Way, $M_@$. From the above formulae we can then see that:
\begin{eqnarray}\label{mclumps}\nonumber
M_{cl}=M(10^{-6}M_{\odot},10^{10}M_{\odot}) & = & \frac{16}{30}M_@\simeq 53.3\,\% M_@     \\  
N_{cl}=N(10^{-6}M_{\odot},10^{10}M_{\odot}) & = & \frac{10^6}{30\ln{10}}\frac{M_@}{M_{\odot}}\simeq 2.90\times10^{17}.  
\end{eqnarray}

\begin{table}[tbp]
\centering
\caption{Density parameters (unbiased case)}
\label{haloparameters}
\begin{tabular}{lccccc}\hline\hline
 	& $r_{0h}$ & $\rho_{0h}$	    & $\rho_{0cl}$           & $c_h$  &   A\\
 	& kpc      & GeV\,$c^{-2}$cm$^{-3}$ & GeV\,$c^{-2}$cm$^{-3}$ &  & $M_{\odot}^{-1}$\,kpc$^{-3}$\\ \hline
NFW	& 21.7	   & 0.132 & 0.151	    & 10.9                      & $1.08\times10^5$ \\
Burkert	& 13.5	   & 0.404 & 0.462	    & 11.5              &   $3.30\times10^5$ \\ \hline\hline
\end{tabular} 
\end{table}
The DM density is the sum of its two components, the smooth halo (h) and he clumpy one (cl):
\[
 \rho_{tot}(r)=\rho_{h}(r)+\rho_{cl}(r).
\]
The spatial distribution is less known, and two different hypotheses have been proposed till now, an \textit{unbiased} distribution which assumes that the subhaloes distribution follows the same radial profile of the main halo, and an \textit{antibiased} distribution which assumes that the two are anti-correlated. We will consider both cases separately.

In the case of unbiased distribution for the substructures we have that $\rho_{h}$ and $\rho_{cl}$ have the same spatial dependence: $\rho_{i}(r) = \rho_{0i} f(r/r_{0})$, in particular they share the same scale parameter $r_{0}$. The three parameters $r_{0}$, $\rho_{0h}$ and $\rho_{0cl}$ describing the DM distribution  can be determined using the three known quantities   $M_{@}$, the total mass of the Milky way, $\rho_S$, the local value of the DM density, and the total mass  in substructures derived in Eq.~(\ref{mclumps}).   
Table \ref{haloparameters}  shows the results obtained assuming 
$M_@=1.49\times10^{12}\,M_{\odot}$, $R_S=8.28$ kpc, $\rho_S=0.389$ GeV\,$c^{-2}$  \cite{Catena:2009}
and two different choices for the DM density. For $M_@$, $R_S$ and $\rho_S$ we use the values obtained from  \cite{Catena:2009} assuming a NFW density profile, because they are determined with a slightly better accuracy. The values obtained under an Einasto hypothesis are compatible with these within one standard deviation \cite{Catena:2009}.

For the anti-biased case we follow the approach of \cite{Pieri:2009je} in which the original antibiased (cored) distribution of subhaloes, initially proposed in \cite{Diemand:2004kx,Diemand:2007qr}, is modified in order to be consistent with a given overall density profile of the Galaxy. The two contributions $\rho_h(r)$ and $\rho_{cl}(r)$ to $\rho_{tot}(r)$ are written in the form
\[
 \rho_h(r)=\frac{1}{1+r/r_b}\rho_{tot}(r)
\]
\vspace{-0.5pc}
\[
\rho_{cl}(r)=\frac{r/r_b}{1+r/r_b}\rho_{tot}(r) 
\]
in terms of the \textit{bias radius} $r_b$, determined by inverting the equation
\[
 \int_0^{r_{\Delta}}\rho_h(r)4\pi r^2 dr = M_h = M_@ - M_{cl}
\]
See \cite{Pieri:2009je} for more details. $r_{0}$ and $\rho_{0}$ are the parameters defining $\rho_{tot} = \rho_0 f(r/r_0)$.
The results are shown in table \ref{haloparameters2}  

\begin{table}[tbp]
\centering
\caption{Density parameters (antibiased case)}
\label{haloparameters2}
\begin{tabular}{lccccc}\hline\hline
 	& $r_{0h}$ & $\rho_{0}$	            & $r_b$ &   A\\
 	& kpc      & GeV\,$c^{-2}$cm$^{-3}$ & kpc & $M_{\odot}^{-1}$\,kpc$^{-3}$\\ \hline
NFW	& 21.7	   & 0.284                  & 61.5  &  $2.03\times10^{5}$ \\
Burkert	& 13.5	   & 0.866                  & 54.4  &  $6.19\times10^{5}$ \\ \hline\hline
\end{tabular} 
\end{table}

The last piece of information required to derive the
annihilation signal from the clumps is  the DM 
distribution within the  clumps themselves. We will assume that the
clumps follow the same mass profile as the main halo, but with their own parameters $r_{0,cl}$ and $\rho _{0,cl}$ replacing the ones associated to the halo. 
We parametrize the internal concentration of the subhaloes like in \cite{Pieri:2009je} (\textit{Via Lactea II} case). In particular, since we deal with the smallest clumps, we fit it to a simple power law in the mass range $10^{-6}\div 10^{4}M_{\odot}$:
\[
c_{{cl}}=c_0
\left(m_{{cl}}/M_{\odot}\right)^{-\beta}
\]
with
$c_0=102.8$ and $\beta=0.0331$.

\section{Mean Electron and positron flux and gradient}
\label{appendix:mean}

\begin{figure}[tbp]
\centering
\includegraphics[width=0.45\textwidth]{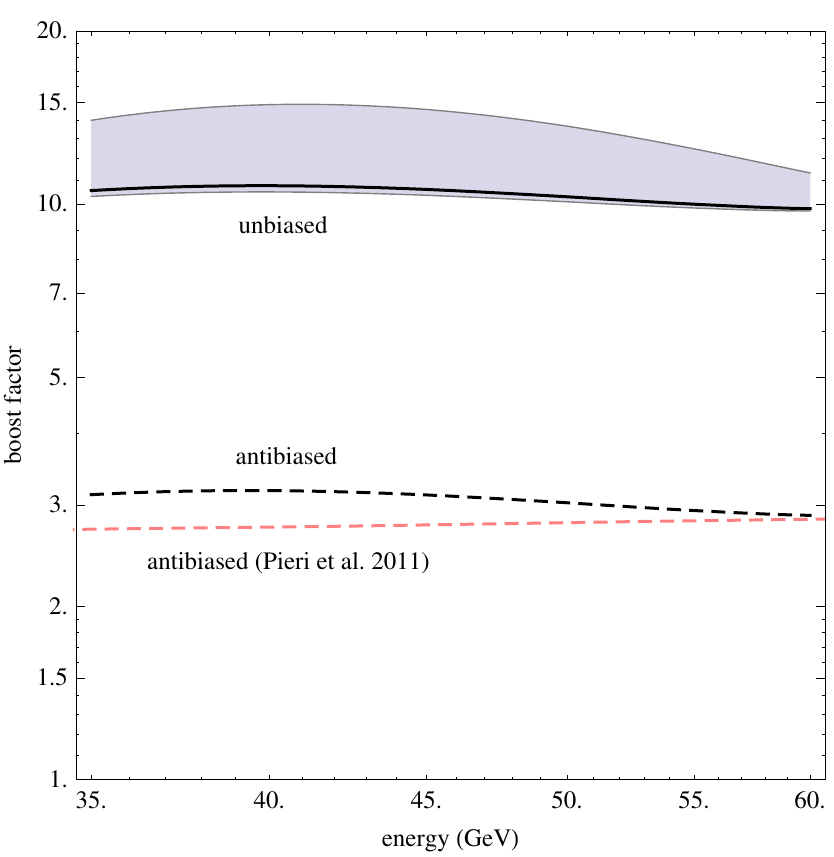}
\caption{Boost factor given by dark matter substructures to the electron and positron fluxes. The solid line shows the case of unbiased DM substructure distribution considered in the paper. The shaded region represents the fluctuations from different realizations of the substructures. The dashed black line corresponds to the antibiased distribution used also in \cite{Pieri:2009je} which is instead shown with the red dashed line.} 
\label{fig:boost} 
\end{figure}
The mean flux from all the substructures can be expressed as:
\begin{eqnarray}
\label{meanflux}
\left\langle\Phi_e (E)\right\rangle &=& \frac{c}{4\pi}\left\langle \sum_{cl} \frac{dn_e}{dE}(E,\textbf{x}_S,m_{cl},\textbf{x}_{cl})\right\rangle\\
&:=& \frac{c}{4\pi}\int \frac{dn_e}{dE}\frac{dn_{cl}}{dm}\,dm\,dV      \nonumber          \\   \nonumber  
&=& \frac{1}{b(E)}\frac{c}{4\pi}\int_E^{m_{\chi} c^2} \frac{\left\langle \sigma_A v \right\rangle}{2} 
    \frac{\left\langle \rho^2_{eff}(E,E',\textbf{x}_S) \right\rangle}{m_{\chi}^2} \frac{dN}{dE'}dE' \ . 
\end{eqnarray}
A similar expression holds for the mean gradient.
The rather complicate equation above actually quite simplifies in the ``far clump'' approximation (see \ref{appendix:diffeq}) where the integration over the clump positions and masses can be performed separately and evaluated in a straightforward way. 

We show in Fig.~\ref{fig:boost} the boost factor of DM substructures to the electron and positron fluxes and we compare it with the same quantity computed in \cite{Pieri:2009je} for a similar DM distribution. The solid line  shows the case of unbiased DM substructure distribution considered in the paper and the shaded region  represent the fluctuations from different realizations of the substructures.  In order to compare our results with those shown in \cite{Pieri:2009je} we also show the case of antibiased substructures' distribution. As it is clear from Fig.~\ref{fig:boost}, passing from an unbiased to an antibiased model reduces the relative contribution by roughly a factor of 3, in very good agreement with the results of \cite{Pieri:2009je}. The residual difference is due to some differences in the global parameters we adopt to describe the dark matter distribution. These differences, however, are not relevant when computing the anisotropy.

\end{document}